\documentclass[preprintnumbers,prd,showpacs,floatfix,superscriptaddress,nofootinbib,twocolumn, letterpaper]{revtex4-1}
\usepackage{longtable}
\usepackage{bm}
\usepackage{relsize}
\usepackage{amsfonts}
\usepackage{amsmath}
\usepackage{amssymb,epsf}
\usepackage{latexsym}
\usepackage{graphicx,epsfig}
\usepackage{amssymb}
\usepackage{float}
\usepackage{subfigure}
\usepackage{epstopdf}
\usepackage[colorlinks=true,citecolor=blue,linkcolor=blue,urlcolor=black]{hyperref}
\usepackage{dcolumn}
\usepackage{psfrag}
\usepackage{wrapfig}
\usepackage{makeidx}
\usepackage{epsf}
\usepackage{color}
\usepackage{multirow}
\usepackage{mathtools}

\begin{document}

\title{Cosmology in scalar-tensor $f(R,T)$ gravity}

\author{Tiago B. Gon\c{c}alves}
\email{tgoncalves@alunos.fc.ul.pt}
\affiliation{Instituto de Astrof\'{i}sica e Ci\^{e}ncias do Espa\c{c}o, Faculdade de Ci\^{e}ncias da Universidade de Lisboa, Edif\'{i}cio C8, Campo Grande, P-1749-016 Lisbon, Portugal}
\affiliation{Departamento de F\'{i}sica, Faculdade de Ci\^{e}ncias da Universidade de Lisboa, Edif\'{i}cio C8, Campo Grande, P-1749-016 Lisbon, Portugal}

\author{Jo\~{a}o Lu\'{i}s Rosa}
\email{joaoluis92@gmail.com}
\affiliation{Institute of Physics, University of Tartu, W. Ostwaldi 1, 50411 Tartu, Estonia}

\author{Francisco S. N. Lobo}
\email{fslobo@fc.ul.pt}
\affiliation{Instituto de Astrof\'{i}sica e Ci\^{e}ncias do Espa\c{c}o, Faculdade de Ci\^{e}ncias da Universidade de Lisboa, Edif\'{i}cio C8, Campo Grande, P-1749-016 Lisbon, Portugal}
\affiliation{Departamento de F\'{i}sica, Faculdade de Ci\^{e}ncias da Universidade de Lisboa, Edif\'{i}cio C8, Campo Grande, P-1749-016 Lisbon, Portugal}
\date{\today}

\begin{abstract} 
In this work, we use reconstruction methods to obtain cosmological solutions in the recently developed scalar-tensor representation of $f(R,T)$ gravity. Assuming that matter is described by an isotropic perfect fluid and the spacetime is homogeneous and isotropic, i.e., the Friedmann-Lema\^{i}tre-Robsertson-Walker (FLRW) universe, the energy density, the pressure, and the scalar field associated with the arbitrary dependency of the action in $T$ can be written generally as functions of the scale factor. We then select three particular forms of the scale factor: an exponential expansion with ${a(t)\propto e^t}$ (motivated by the de Sitter solution); and two types of power-law expansion with ${a(t)\propto t^{1/2}}$ and ${a(t)\propto t^{2/3}}$ (motivated by the behaviors of radiation- and matter-dominated universes in general relativity, respectively). A complete analysis for different curvature parameters ${k=\{-1,0,1\}}$ and equation of state parameters ${w=\{-1,0,1/3\}}$ is provided. Finally, the explicit forms of the functions $f\left(R,T\right)$ associated with the scalar-field potentials of the representation used are deduced.
\end{abstract}


\maketitle

\section{Introduction}\label{sec:intro}

A plethora of gravitational theories related to extensions of Einstein's theory of general relativity 
(GR) have been proposed in the literature (see, e.g., Refs.~\cite{Nojiri:2006ri,Lobo:2008sg,Nojiri:2010wj,Clifton:2011jh,Capozziello:2011et,CANTATA:2021ktz,Avelino:2016lpj}), 
essentially motivated by the potential to explain the late-time accelerated expansion of the universe 
\cite{SupernovaCosmologyProject:1998vns,SupernovaSearchTeam:1998fmf}. One of the simplest 
modifications of GR is the so called $f(R)$ gravity \cite{Sotiriou:2008rp}, in which the gravitational Lagrangian instead of depending linearly on the Ricci curvature scalar $R$ is allowed to depend on a general function of $R$. Giving rise to modified Friedmann equations, $f(R)$ gravity can indeed be consistent with an accelerated expansion without necessarily requiring a dark energy component \cite{Capozziello:2002rd}. At a more fundamental level, one may consider different approaches to $f(R)$ gravity, namely, the metric formalism which consists in varying the action with respect to the metric \cite{Sotiriou:2008rp}, the metric-affine formalism where the metric and the connections are treated as separate variables \cite{Olmo:2011uz} and the hybrid formalism \cite{Harko:2011nh,Harko:2020ibn,Harko:2018ayt,Capozziello:2012ny,Capozziello:2013uya,Capozziello:2015lza,Rosa:2017jld,Rosa:2019ejh,Rosa:2021ish} which unifies the above-mentioned approaches.

An option is to explore other extensions analogous to $f(R)$, for instance, $f(R,T)$ gravity where now the gravitational Lagrangian is allowed to depend on a general function not only of $R$ but also of the trace of the stress-energy tensor $T$ \cite{Harko:2011kv}. Note that the inclusion of the $T$-dependence may be due to quantum effects, such as conformal anomalies, or may also be induced by relativistically covariant models of interacting dark energy, where a cosmological term in the gravitational Lagrangian is a function of the trace of the stress-energy, $\Lambda(T)$ \cite{Poplawski:2006ey}. It is interesting to note that in these types of theories the matter stress-energy tensor is generally not conserved, due to the explicit coupling between the geometry and matter, and this non-conservation determines the appearance of an extra-force acting on the particles in motion in the gravitational field. In fact, the equations of motion of the test particles are non-geodesic, and take place in the presence of an extra-force orthogonal to the four-velocity. The astrophysical and cosmological applications of $f(R,T)$ gravity have received an extensive attention in the literature (we refer the reader to \cite{Harko:2018ayt} for more details).

For instance, relative to the cosmological applications, one may mention reconstruction methods of several cosmological models \cite{Jamil:2011ptc}, in particular, from holographic dark energy \cite{Houndjo:2011fb} and of the matter dominated and accelerated phases \cite{Houndjo:2011tu}, the evolution of scalar cosmological perturbations \cite{Alvarenga:2013syu}, dynamical system approaches were also explored \cite{Shabani:2013djy,Shabani:2014xvi}, applications to 5-D were also explored \cite{Moraes:2015kka} and in particular, solutions of thick branes in the scalar-tensor representation of $f(R,T)$ gravity were found \cite{Rosa:2021tei,Rosa:2021myu}, among many other applications. An interesting generalization of $f(R,T)$ gravity is the inclusion of a contraction $R_{\mu\nu}T^{\mu\nu}$ \cite{Haghani:2013oma,Odintsov:2013iba}, as for the specific case of $T=0$ reduces to $f(R)$ gravity. The cosmological applications of these theories have been extensively explored in Refs.~\cite{Haghani:2013oma,Odintsov:2013iba}.
Relative to the astrophysical applications, much work has been explored in the literature, such as in application to  dark matter \cite{Zaregonbadi:2016xna} and the study of compact objects  \cite{Moraes:2017mir,Zubair:2016cde,Moraes:2016akv,Das:2016mxq}

Often, these modified theories of gravity can be rewritten in an equivalent scalar-tensor representation. Scalar-tensor theories of gravity have been extensively studied and they are useful in modeling deviations from GR. One of advantages is the relative simplicity of their field equations which allows analytical solutions to be found in various physical systems \cite{Clifton:2011jh}. In fact, given the large number of models, the question that arises consists on how one could study and compare them in a unified manner and, in particular, determine which if any is the origin of cosmic acceleration. Indeed, a particularly useful tool in this direction, was the discovery that these classes of models are specific cases of the most general Lagrangian which
leads to second order field equations, namely, the Horndeski Lagrangian \cite{Horndeski:1974wa}, which was recently rediscovered \cite{Deffayet:2011gz}. This realization enables one to adopt a unifying framework, and to determine subsets within this general theory that have appealing theoretical properties.
Theories for which this has been done extensively are, e.g., $f(R)$ gravity in the metric and the Palatini formalism, and hybrid metric-Palatini gravity \cite{Harko:2018ayt}. Yet, only recently has a scalar-tensor representation in $f(R,T)$ gravity been proposed \cite{Rosa:2021teg}, which was used to study junction conditions for the matching between two spacetimes at a separation hypersurface, and to find thick brane solutions \cite{Rosa:2021tei,Rosa:2021myu}. Otherwise, the scalar-tensor representation of $f(R,T)$ gravity remains largely unexplored.

In this work, we explore cosmological solutions in the scalar-tensor representation of $f(R,T)$ gravity. In fact, the interest in studying modifications of GR in a cosmological setting comes from the extra degrees of freedom that become available in these theories. For instance, the scale factor is no longer uniquely determined by the matter content of the universe. In other words, one can choose a scale factor motivated by observations, such as, an exponential or power law, and verify if the theory provides solutions that are consistent with different contributions to the energy density, e.g., matter, radiation or cosmological constant. Therefore, an exponential expansion, for instance, might no longer require dark energy. As opposed to specifying a form of the action \textit{a priori} in order to find what solutions it may lead to, which is often a more complicated procedure; this process of starting from the observed evolution of the universe in order to find consistent solutions and recover the form of the gravitational action is known as reconstruction. Reconstruction methods have been used with other modified theories of gravity such as $f(R)$ gravity \cite{Capozziello:2005ku,Multamaki:2005zs}. In the case of $f(R,T)$, there are two extra degrees of freedom, when compared to GR, which corresponds to two scalar fields in its scalar-tensor representation. Thus, the focus of this work is to derive the equations of motion and use reconstruction methods in scalar-tensor $f(R,T)$ gravity. 

The structure of this paper is as follows. In Sec.~\ref{sec:fRTintro} we introduce both the original geometrical representation and the equivalent scalar-tensor representation of $f(R,T)$ gravity. In Sec.~\ref{sec:cosmo} we assume a Friedmann-Lema\^{i}tre-Robertson-Walker (FLRW) universe filled with an isotropic and homogeneous perfect fluid, whose stress-energy tensor we require that it be conserved. For these assumptions, we show the derived system of equations of motion and obtain a partial solution to the system before imposing any further constraints. These we impose in Sec.~\ref{sec:particular} finding complete solutions by choosing different forms of the scale factor and the values of the curvature parameter and equation of state. Sec.~\ref{sec:fRT-explicit} is where we present our attempts to find explicit forms of the function $f(R,T)$ in the particular cases we consider. Finally, the discussion of our results and future prospects are in Sec.~\ref{sec:conclusion}.

\section{Theory and equations of the $f\left(R,T\right)$ gravity}\label{sec:fRTintro}

\subsection{Geometrical representation}

The $f\left(R,T\right)$ gravity theory is described by the following action \cite{Harko:2011kv}
\begin{equation}\label{eq:fRTaction-original}
    S = \frac{1}{2\kappa^2} \int_{\Omega}\sqrt{-g} \, f(R,T) d^4 x+ \int_{\Omega} \sqrt{-g} \, \mathcal{L}_m d^4 x,
\end{equation}
where $\kappa^2=8\pi G/c^4$, $G$ is the gravitational constant and $c$ is the speed of light, $\Omega$ is the 4-dimensional spacetime manifold on which one defines a set of coordinates $x^\mu$, $g$ is the determinant of the metric $g_{\mu\nu}$, $f\left(R,T\right)$ is an arbitrary well-behaved function of the Ricci scalar $R=g^{\mu\nu}R_{\mu\nu}$, where $R_{\mu\nu}$ is the Ricci tensor, and the trace $T=g^{\mu\nu}T_{\mu\nu}$ of the stress-energy tensor $T_{\mu\nu}$. The latter is defined in terms of the variation of the matter Lagrangian $\mathcal L_m$ as
\begin{equation}
T_{\mu\nu}=-\frac{2}{\sqrt{-g}}\frac{\delta\left(\sqrt{-g}\mathcal L_m\right)}{\delta g^{\mu\nu}}.
\end{equation}
In the following, we adopt a system of geometrized units in such a way that $G=c=1$, and thus $\kappa^2=8\pi$. 

Taking the variation of Eq.~\eqref{eq:fRTaction-original} with respect to the metric $g_{\mu\nu}$ leads to the field equations of $f(R,T)$ gravity (see Ref.~\cite{Harko:2011kv} for details)
\begin{equation}\label{eq:fields-original}
\begin{multlined}
    f_R R_{\mu\nu}-\frac{1}{2}g_{\mu\nu}f(R,T) + \left(g_{\mu\nu}\square-\nabla_\mu\nabla_\nu\right)f_R \\ 
    = \kappa^2 T_{\mu\nu}-f_T (T_{\mu\nu}+\Theta_{\mu\nu}),
\end{multlined}
\end{equation}
where the subscripts $R$ and $T$ denote partial derivatives of $f$ with respect to these variables, respectively, $\nabla_\mu$ is the covariant derivative and $\square\equiv\nabla^\sigma\nabla_\sigma$ is the D'Alembert operator, both defined in terms of the metric $g_{\mu\nu}$, and the tensor $\Theta_{\mu\nu}$ is defined as
\begin{equation}\label{eq:Theta-varT}
    \Theta_{\mu\nu}\equiv g^{\rho\sigma}\frac{\delta T_{\rho\sigma}}{\delta g^{\mu\nu}}.
\end{equation}

The conservation equation for $f\left(R,T\right)$ gravity can be obtained by taking the divergence of Eq.~\eqref{eq:fields-original} and using the identity $\left(\square\nabla_\nu-\nabla_\nu\square\right)f_R=R_{\mu\nu}\nabla^\mu f_R$, from which one obtains
\begin{equation}\label{eq:conserv-general}
\begin{multlined}
    (\kappa^2-f_T)\nabla^\mu T_{\mu\nu}=\left(T_{\mu\nu}+\Theta_{\mu\nu}\right)\nabla^\mu f_T \\ 
    +f_T\nabla^\mu\Theta_{\mu\nu}+f_R \nabla^\mu R_{\mu\nu}-\frac{1}{2}g_{\mu\nu}\nabla^\mu f.
\end{multlined}
\end{equation}

We next consider the scalar-tensor representation of $f(R,T)$ gravity, which will be used throughout this work.

\subsection{Scalar-tensor representation}\label{subsec:scalar-tensor}

Similarly to other modified theories of gravity featuring extra scalar degrees of freedom in comparison to GR, one can deduce a dynamically equivalent scalar-tensor representation of $f\left(R,T\right)$ gravity with two scalar fields. To do so, one introduces two auxiliary fields $\alpha$ and $\beta$ and rewrites the action \eqref{eq:fRTaction-original} in the form 
\begin{eqnarray}\label{eq:STaction-intro}
    S &=& \frac{1}{2\kappa^2} \int_{\Omega}\sqrt{-g} \big[f(\alpha,\beta)+ (R-\alpha)f_\alpha
		\nonumber \\     
    && +(T-\beta)f_\beta\big] d^4 x  + \int_{\Omega} \sqrt{-g}\mathcal{L}_m  d^4 x , 
\end{eqnarray}
where the subscripts $\alpha$ and $\beta$ denote partial derivatives with respect to these variables, respectively. Taking the variation of Eq.~\eqref{eq:STaction-intro} with respect to $\alpha$ and $\beta$, the equations of motion for the fields $\alpha$ and $\beta$ are found to be, respectively, 
\begin{eqnarray}
    (R-\alpha)f_{\alpha\alpha}+(T-\beta)f_{\alpha\beta} &=& 0, \label{eq:motion-alpha} \\
    (R-\alpha)f_{\beta\alpha}+(T-\beta)f_{\beta\beta} &=& 0. \label{eq:motion-beta}
\end{eqnarray}
The two equations above can be rewritten in a matrix form $\mathcal M\textbf{x}=0$ as
\begin{equation}
\begin{pmatrix}
f_{\alpha\alpha} & f_{\alpha\beta} \\
f_{\beta\alpha} & f_{\beta\beta}
\end{pmatrix}
\begin{pmatrix}
R-\alpha\\
T-\beta
\end{pmatrix}=0.
\end{equation}
Matrix equations of this form are known to yield a unique solution if and only if the determinant of the matrix $\mathcal M$ is non-vanishing, i.e., $f_{\alpha\alpha}f_{\beta\beta}\neq f_{\alpha\beta}^2$. In such a case, the unique solution is $R=\alpha$ and $T=\beta$. Inserting these results back into Eq.~\eqref{eq:STaction-intro}, one verifies that this equation reduces to the form of action \eqref{eq:fRTaction-original}, proving the equivalence between the two representations, and the scalar-tensor representation is well defined.

Defining two scalar fields $\varphi$ and $\psi$ and a scalar interaction potential $V\left(\varphi,\psi\right)$ in the forms
 \begin{equation}\label{eq:varphi&psi}
     \varphi\equiv\frac{\partial f}{\partial R} ,\qquad
    \psi\equiv\frac{\partial f}{\partial T},
 \end{equation}
\begin{equation}\label{eq:potential}
    V(\varphi,\psi) \equiv -f(\alpha,\beta)+ \varphi \alpha + \psi \beta ,
\end{equation}
one can rewrite action \eqref{eq:STaction-intro} in the equivalent scalar-tensor representation as
\begin{equation}\label{eq:STaction}
    \begin{split} 
    S = \frac{1}{2\kappa^2} \int_{\Omega} \sqrt{-g} \left[\varphi R+\psi T - V(\varphi, \psi)\right]d^4 x \\ 
    + \int_{\Omega} \sqrt{-g} \mathcal{L}_m d^4 x . 
    \end{split}
\end{equation}
Similarly to what happens in the metric representation of $f(R)$ theories of gravity, the scalar field $\varphi$ is analogous to a Brans-Dicke scalar field with parameter $\omega_{BD}=0$ and with an interaction potential $V$. In addition to this scalar field $\varphi$, the second scalar degree of freedom of $f(R,T)$ gravity, associated with the arbitrary dependence of the action in $T$, is also represented by a scalar field, $\psi$.

The action \eqref{eq:STaction} depends on three independent variables, the metric $g_{\mu\nu}$ and the two scalar fields $\varphi$ and $\psi$. Varying this action with respect to the metric $g_{\mu\nu}$ yields the field equations
\begin{equation}\label{eq:fields}
    \begin{multlined}
      \varphi R_{\mu\nu}-\frac{1}{2}g_{\mu\nu}\left(\varphi R + \psi T - V\right)\\-(\nabla_\mu\nabla_\nu-g_{\mu\nu}\square)\varphi = \kappa^2 T_{\mu\nu} -\psi (T_{\mu\nu} + \Theta_{\mu\nu}),
      \end{multlined}
\end{equation}
which could also be obtained directly from Eq.~\eqref{eq:fields-original} by using the definitions shown in Eqs.~\eqref{eq:varphi&psi} and \eqref{eq:potential}, with $\alpha=R$ and $\beta=T$. Furthermore, taking the variation of Eq.~\eqref{eq:STaction} with respect to the scalar fields $\varphi$ and $\psi$ gives, respectively,
\begin{equation}\label{eq:Vphi}
    V_{\varphi} = R,
\end{equation}
\begin{equation}\label{eq:Vpsi}
    V_{\psi} = T,
\end{equation}
where the subscripts in $V_\varphi$ and $V_\psi$ denote the derivatives of the potential $V(\varphi,\psi)$ with respect to the variables $\varphi$ and $\psi$, respectively.\par

Additionally, using the same definitions [Eqs.~\eqref{eq:varphi&psi} and \eqref{eq:potential}, with $\alpha=R$ and $\beta=T$] and the geometrical result $\nabla^\mu\left(R_{\mu\nu}-\frac{1}{2}g_{\mu\nu}R\right)=0$, the conservation equation for $f\left(R,T\right)$ gravity in the scalar-tensor representation becomes
\begin{equation}\label{eq:conserv-general2}
\begin{multlined}
    (\kappa^2-\psi)\nabla^\mu T_{\mu\nu}=\left(T_{\mu\nu}+\Theta_{\mu\nu}\right)\nabla^\mu\psi+ \\ +\psi\nabla^\mu\Theta_{\mu\nu}-\frac{1}{2}g_{\mu\nu}\left[R\nabla^\mu\varphi+\nabla^\mu\left(\psi T-V\right)\right].
\end{multlined}
\end{equation}

We will use the results obtained in this subsection to explore cosmological solutions in the next section.

\section{Cosmological Equations}\label{sec:cosmo}

\subsection{Framework and assumptions}

In this work, we assume that the universe is well-described by an homogeneous and isotropic FLRW spacetime, which in the usual spherical coordinates $(t,r,\theta,\phi)$ takes the form
\begin{equation}\label{eq:FLRW-metric}
    ds^2 = -dt^2+a^2(t)\left[\frac{dr^2}{1-kr^2}+r^2\left(d\theta^2+\sin^2\theta d\phi^2\right)\right], 
\end{equation}
where $a(t)$ is the scale factor and $k$ is the curvature parameter which can take the values $k=\left\{-1,0,1\right\}$ corresponding to a hyperbolic, spatially flat, or hyperspherical universe, respectively.\par

We also assume that matter is described by an isotropic perfect fluid, i.e., the stress-energy tensor $T_{\mu\nu}$ is given by
\begin{equation}\label{eq:em-fluid}
    T_{\mu\nu}=(\rho+p)u_\mu u_\nu +p g_{\mu\nu},
\end{equation}
where $\rho$ is the energy density, $p$ is the isotropic pressure, and $u^\mu$ is the fluid 4-velocity satisfying the normalization condition $u_\mu u^\mu=-1$. Taking the matter Lagrangian to be $\mathcal L_m=p$ \cite{Bertolami:2008ab}, the tensor $\Theta_{\mu\nu}$ takes the form
\begin{equation}\label{eq:Theta-fluid}
    \Theta_{\mu\nu}=-2T_{\mu\nu}+p g_{\mu\nu}.
\end{equation}

To preserve the homogeneity and isotropy of the solution, all physical quantities are assumed to depend solely on the time coordinate $t$, i.e., $\rho=\rho\left(t\right)$, $p=p\left(t\right)$, $\varphi=\varphi\left(t\right)$, and $\psi=\psi\left(t\right)$. Under these assumptions, one obtains two independent field equations from Eq.~\eqref{eq:fields}, namely, the modified Friedmann equation and the modified Raychaudhuri equation, which take the following forms
\begin{equation}\label{eq:tt}
    \dot{\varphi}\left(\frac{\dot{a}}{a}\right) + \varphi \left( \frac{\dot{a}^{2} + k}{a^{2}} \right)  = \frac{8 \pi}{3}\rho + \frac{\psi}{2}\left( \rho - \frac{1}{3}p\right)+ \frac{1}{6} V,
\end{equation}
\begin{equation}\label{eq:rr}
\begin{split}
   \ddot{\varphi}+ 2\dot{\varphi}\left(\frac{\dot{a}}{a}\right) +\varphi\left( \frac{2\ddot{a}}{a} + \frac{\dot{a}^{2}+k}{a^{2}} \right)  = -8\pi p  \\+ \frac{\psi}{2}\left(\rho-3p\right) + \frac{1}{2} V,
\end{split}
\end{equation}
respectively, where overdots denote derivatives with respect to time. Furthermore, the equations of motion for the scalar fields $\varphi$ and $\psi$ from Eqs.~\eqref{eq:Vphi} and \eqref{eq:Vpsi} become
\begin{equation}\label{eq:Vphicosmo}
    V_{\varphi} =R= 6\left( \frac{\ddot{a}}{a}+ \frac{\dot{a}^{2}+k}{a^{2}}\right),
\end{equation}
\begin{equation}\label{eq:Vpsicosmo}
    V_{\psi} =T= 3p-\rho,
\end{equation}
respectively. Finally, the conservation equation from Eq.~\eqref{eq:conserv-general2} in this framework takes the form
\begin{equation}\label{eq:conserv-total}
\begin{multlined}
    8\pi(\rho+p)\left(\frac{\dot{a}}{a}\right)+\frac{8\pi}{3}\dot{\rho} = \dot{\varphi}\left(\frac{\ddot{a}}{a}+\frac{\dot{a}^2+k}{a^2}-\frac{1}{6}V_\varphi\right) \\
    -\dot{\psi}\left(\frac{1}{2}\rho - \frac{1}{6}p+\frac{1}{6}V_\psi\right) -\psi\left[\frac{\dot{a}}{a}(\rho+p)+\frac{1}{2}\dot{\rho} - \frac{1}{6}\dot{p}\right].
\end{multlined}
\end{equation}

The system of Eqs.~\eqref{eq:tt}--\eqref{eq:conserv-total} forms a system of five equations from which only four are linearly independent. To prove this feature, one can take the time derivative of Eq.~\eqref{eq:tt}, use Eqs.~\eqref{eq:Vphicosmo} and \eqref{eq:Vpsicosmo} to eliminate the partial derivatives $V_\varphi$ and $V_\psi$, use the conservation equation in Eq.~\eqref{eq:conserv-total} to eliminate the time derivative $\dot\rho$, and use the Raychaudhuri equation in Eq.~\eqref{eq:rr} to eliminate the second time derivative $\ddot a$, thus recovering the original equation. Thus, one of these equations can be discarded from the system without loss of generality. Given the complicated nature of Eq.~\eqref{eq:rr}, we chose to discard this equation and consider only Eqs.~\eqref{eq:tt}, \eqref{eq:Vphicosmo}, \eqref{eq:Vpsicosmo}, and \eqref{eq:conserv-total}.

Note that, in general, modified theories of gravity with geometry-matter couplings \cite{Harko:2011kv,Koivisto:2005yk,Bertolami:2007gv,Harko:2010mv,Harko:2012hm,Haghani:2013oma,Harko:2014gwa,Harko:2014sja,Harko:2014aja,Avelino:2016lpj,Harko:2018gxr,Harko:2020ibn} imply the non-conservation of the matter stress-energy tensor $\nabla_\mu T^{\mu\nu} \neq 0$, which may entail a transfer of energy from the geometry to the matter sector \cite{Harko:2014pqa,Harko:2015pma,Harko:2018ayt,Harko:2021bdi}. Nevertheless, for simplicity, in this work we impose the conservation of the matter stress-energy tensor, and thus obtain the usual cosmological conservation equation of the form 
\begin{equation}\label{eq:conserv-m}
    \dot{\rho} = -3\frac{\dot{a}}{a}(\rho+p).
\end{equation}
Inserting this result back into the conservation equation in Eq.~\eqref{eq:conserv-total} and using Eqs.~\eqref{eq:Vphicosmo} and \eqref{eq:Vpsicosmo} to cancel the factors $V_\varphi$ and $V_\psi$, one obtains a simplified form of the general conservation equation as
\begin{equation}\label{eq:conserv-psi}
       2\dot{\psi}(\rho+p) = -\psi(\dot{\rho}-\dot{p}).
\end{equation} 

Finally, we impose an equation of state of the form
\begin{equation}\label{eq:state}
p=w\rho,
\end{equation}
where $w$ is a dimensionless parameter. We are thus left with a system of six independent equations, namely Eqs.~\eqref{eq:tt}, \eqref{eq:Vphicosmo}, \eqref{eq:Vpsicosmo}, \eqref{eq:conserv-m}, \eqref{eq:conserv-psi}, and \eqref{eq:state}, for the eight independent unknowns $a$, $\rho$, $p$, $\varphi$, $\psi$, $V$, $k$, $w$. Since $V$ is a function of two variables, $\varphi$ and $\psi$, this unknown effectively contributes with two degrees of freedom, and thus one is left with six equations for a total of nine degrees of freedom. Consequently, one can impose up to three constraints to close the system and obtain the solutions, provided that none of these constraints is an explicit form of $V$.

\subsection{General solutions}\label{subsec:gensol}

Despite the under determination of the system of equations derived in the previous section, there are still some general relations between unknowns obtainable without determining the system. In this section, we will derive these relations in the most possible general way, i.e., without imposing any constraints to the system.

\begin{figure}
	\includegraphics[width=0.9\columnwidth]{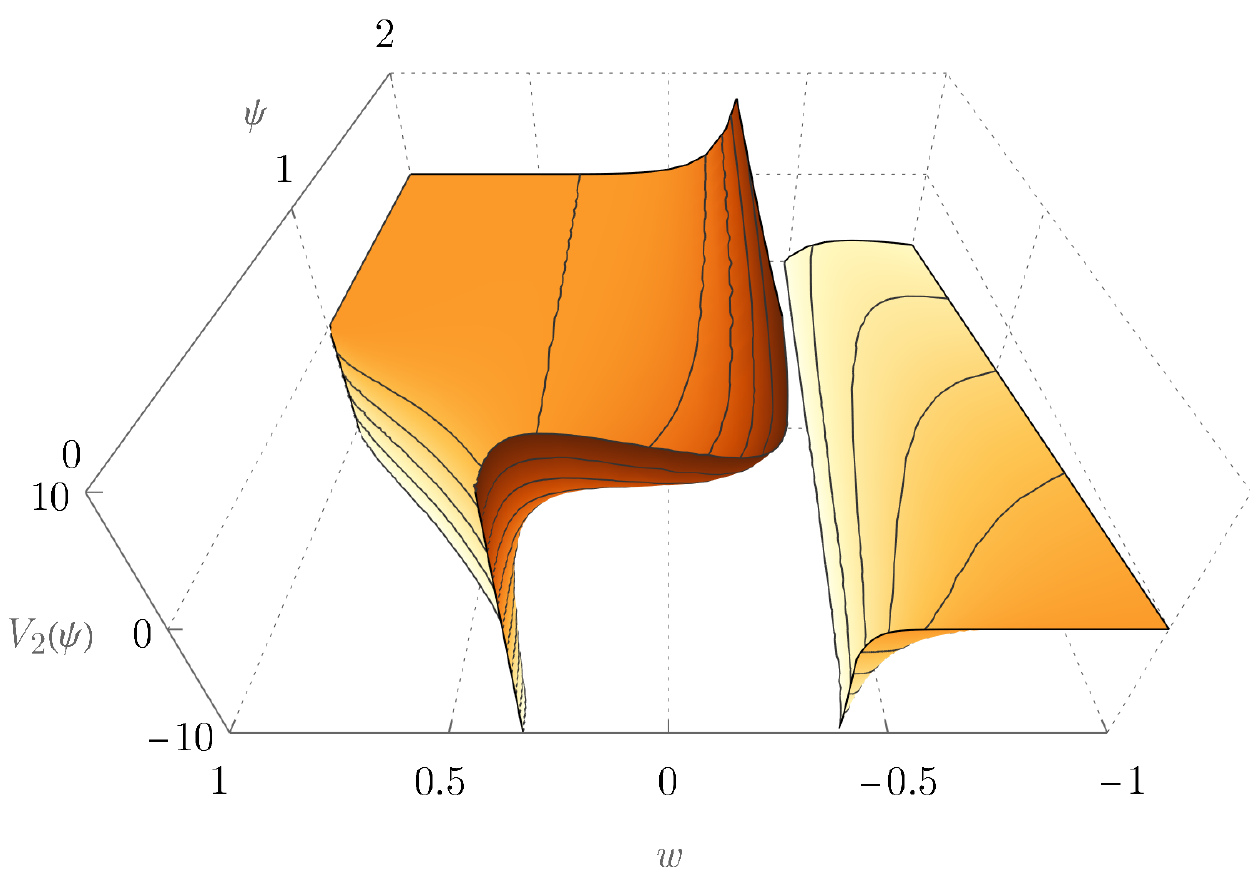}
	\caption{ Visualisation of $V_2(\psi)$ as given by Eq.~\eqref{eq:solV2_psi}, for a range of values $0<\psi<2$ and $-1<w<1$, setting ${\rho_0=\psi_0=1}$. The curves on the graph are contour lines. This function is undefined when $w=\left\{-1/3,1\right\}$ and at $\psi=0$ when $-1/3<w<1$. $V_2(\psi)=0$ when $w=1/3$ ($T=0$).}
	\label{fig:Plot3D_V2_psi}
\end{figure}

Similarly to what happens in GR, the conservation equation for the matter sector given in Eq.~\eqref{eq:conserv-m} associated with the equation of state in Eq.~\eqref{eq:state} allows one to perform an integration and obtain the general forms of the energy density $\rho$ and pressure $p$ as functions of the scale factor as
\begin{equation}\label{eq:solrho}
    \rho = \rho_0 \left(\frac{a}{a_0} \right)^{-3(1+w)}, 
\end{equation}
\begin{equation}\label{eq:solp}
p = w \rho_0 \left(\frac{a}{a_0} \right)^{-3(1+w)},
\end{equation}
where $\rho_0$ and $a_0$ are arbitrary integration constants, where the subscript $0$ denotes the value of the quantity at some instant $t=t_0$ representing e.g. the present time. One can now introduce these explicit forms of $\rho$ and $p$ into the simplified conservation equation in Eq.~\eqref{eq:conserv-psi}, and perform an integration to obtain an explicit form of $\psi$ as a function of $a$ as
\begin{equation}\label{eq:solpsi}
    \psi=\psi_0 \left(\frac{a}{a_0}\right)^{\frac{3}{2}\left(1-w\right)},
\end{equation}
where $\psi_0$ is an integration constant. Notice that (for $w\neq1$) Eq.~\eqref{eq:solpsi} is invertible, i.e. one can write $a(\psi)$. Now, inserting the solutions from Eqs.~\eqref{eq:solrho} and \eqref{eq:solp} into Eq.~\eqref{eq:Vpsicosmo} and using the inverse of Eq.~\eqref{eq:solpsi} to write $a$ as a function of $\psi$, one obtains an explicit form of $V_\psi$ as a function of $\psi$ as
\begin{equation}\label{eq:solVpsi-psi}
	V_\psi = (3w-1)\rho_0\left(\frac{\psi}{\psi_0}\right)^{-\frac{2\left(1+w\right)}{\left(1-w\right)}},
\end{equation}
which is undefined at $w=1$. Since $V_\psi$ does not depend explicitly on the scalar field $\varphi$, this implies that the potential $V\left(\varphi,\psi\right)$, assumed to be arbitrary, is in fact separable in the variables $\varphi$ and $\psi$, i.e., one can write $V\left(\varphi,\psi\right)=V_0+V_1\left(\varphi\right)+V_2\left(\psi\right)$, where $V_0$ is an arbitrary constant and the functions $V_1(\varphi)$ and $V_2(\psi)$ each depends solely on one scalar field, $\varphi$ and $\psi$ respectively. Integrating Eq.~\eqref{eq:solVpsi-psi} with respect to $\psi$ yields
\begin{equation}\label{eq:solV2_psi}
V_2(\psi)=\frac{(1-3w)(1-w)}{(1+3w)}\rho_0 \psi_0^{\frac{2(1+w)}{(1-w)}}\psi^{-\frac{(1+3w)}{(1-w)}},
\end{equation}
and any constant term from integration can be absorbed into $V_0$. This dependence of the potential on $\psi$ vanishes for $w=1/3$ (corresponding to the case when $T=0$) and diverges near $w=\left\{-1/3,1\right\}$. The exponent of $\psi$ is negative for $-1/3<w<1$, so, in this range, $V_2(\psi)$ diverges near $\psi=0$. $V_2(\psi)$ is shown in Fig.~\ref{fig:Plot3D_V2_psi}, for a range of values of the equation of state, where the discontinuities are visible.

At this point, only Eqs.~\eqref{eq:tt} and \eqref{eq:Vphicosmo} remain to be solved. These equations can now be written in the following forms
\begin{equation}\label{eq:ttsub}
\begin{multlined}
    \dot{\varphi}\left(\frac{\dot{a}}{a}\right) + \varphi \left( \frac{\dot{a}^{2} + k}{a^{2}} \right)  = \frac{8 \pi}{3}\rho_0 \left(\frac{a}{a_0}\right)^{-3(1+w)} \\
    +\frac{2(1+w)}{3(1+3w)}\rho_0\psi_0 \left(\frac{a}{a_0}\right)^{-\frac{3}{2}\left(1+3w\right)} + \frac{1}{6} \left[V_0+V_1(\varphi)\right],
\end{multlined}
\end{equation}
\begin{equation}\label{eq:dV1}
\frac{d V_1}{d\varphi} = 6\left( \frac{\ddot{a}}{a}+ \frac{\dot{a}^{2}+k}{a^{2}}\right),
\end{equation}
respectively. 

As it will become useful later on, an equation relating solely $\varphi$ and $a$ can be obtained by taking the derivative of Eq.~\eqref{eq:ttsub} with respect to time, using the chain rule on $V_1$, and using Eq.~\eqref{eq:dV1} to eliminate the derivative of $V_1$ with respect to $\varphi$, from which we obtain
\begin{equation}\label{eq:dttsub}
	 \begin{multlined}
	 \ddot{\varphi} -\dot{\varphi}\left(\frac{\dot{a}}{a} \right) +2\varphi\left(\frac{\ddot{a}}{a}-\frac{\dot{a}^2+k}{a^2}\right)  
=  -\left(1+w\right) \rho_0  \\
 \times	  \left[8\pi \left(\frac{a}{a_0}\right)^{-3(1+w)} + \psi_0 \left(\frac{a}{a_0}\right)^{-\frac{3}{2}\left(1+3w\right)}\right],
	 \end{multlined}
\end{equation}
Two of these equations are independent an they are to be solved for a total of five unknowns, $\varphi$, $a$, $V_1$, $w$ and $k$. In the following, the unknowns $k$ and $w$ will be dealt with as free parameters, at times choosing the particular values $k=\{-1,0,1\}$ and $w=\{-1,0,1/3\}$, where the latter values represent the equations of state for constant dark energy (cosmological constant), dust, and radiation, respectively. To proceed further, an explicit form of $a\left(t\right)$ has to be imposed to determine the system.

\section{Particular Solutions}\label{sec:particular}

\subsection{Exponential expansion with $a\propto e^{t}$}\label{subsec:dS}

\begin{figure}
	\includegraphics[width=0.9\columnwidth]{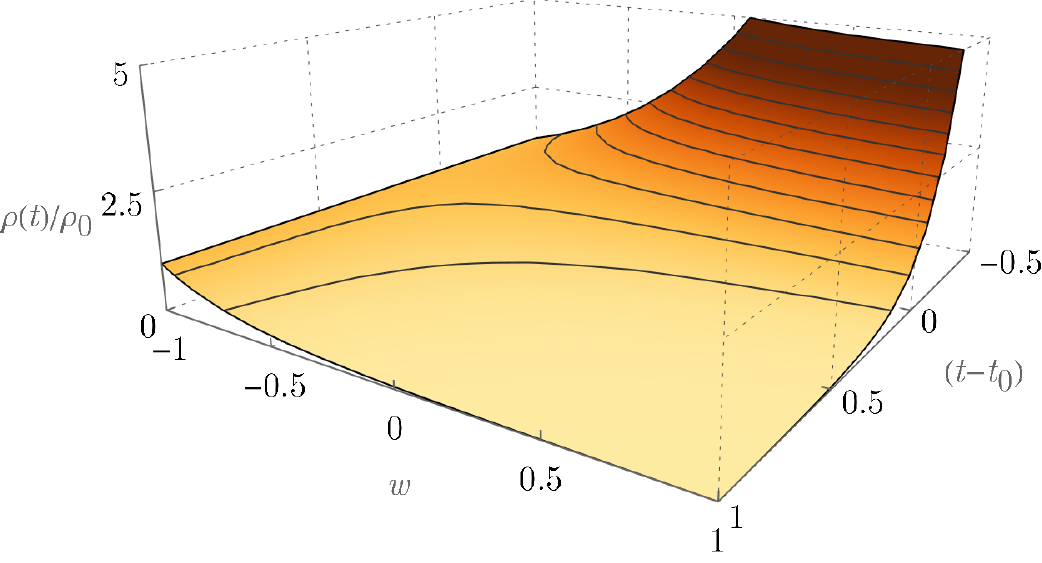}
	\caption{Energy density $\rho\left(t\right)$ from Eq.~\eqref{eq:solrho-dS} in the particular case where $a\propto e^t$ as given by Eq.~\eqref{eq:a-dS} for a range of values of the equation of state parameter, with $\Lambda=1$.}
	\label{fig:Plot_dS_Arho}
\end{figure}

\begin{figure}
	\includegraphics[width=0.9\columnwidth]{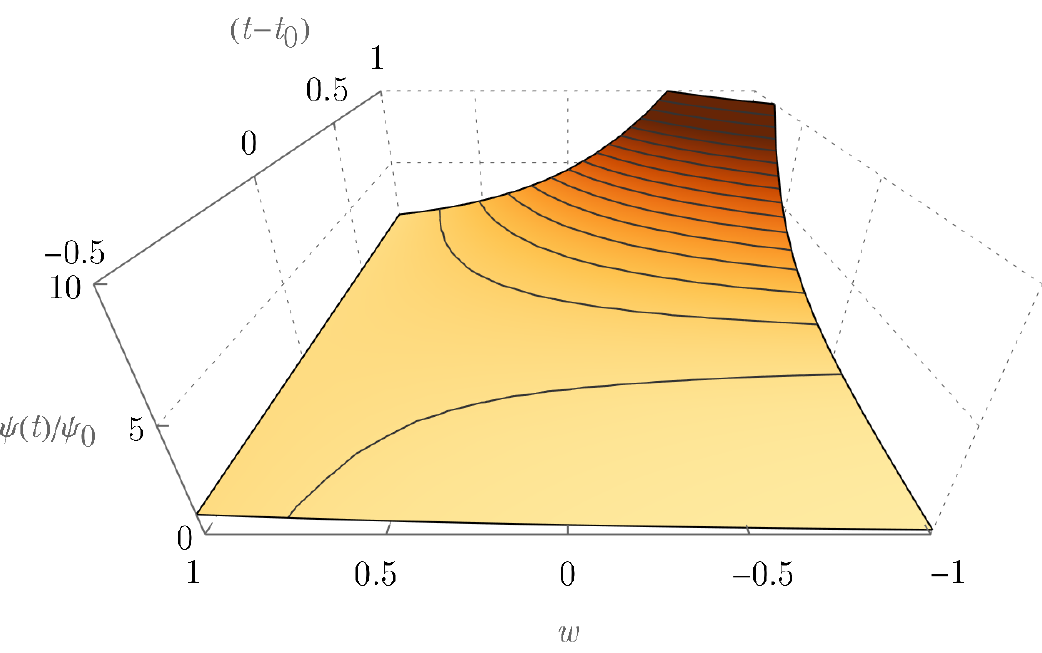}
	\caption{Scalar field $\psi\left(t\right)$ from Eq.~\eqref{eq:solpsi-dS} in the particular case where $a\propto e^t$ as given by Eq.~\eqref{eq:a-dS} for a range of values of the equation of state parameter, with $\Lambda=1$.}
	\label{fig:Plot_dS_Apsi}
\end{figure}

The first particular form of the scale factor we analyze is the de Sitter solution, i.e., an exponentially accelerated expansion. The scale factor takes the form
\begin{equation}\label{eq:a-dS}
    a(t)=a_0 e^{\sqrt{\Lambda}\left(t-t_0\right)},
\end{equation}
where $a_0$, $t_0$ and $\Lambda$ are constants. The constant $a_0$ denotes the value of the scale factor at the time $t_0$, which can be taken to be the present cosmological time, for instance, whereas the constant $\Lambda$ plays the role of a cosmological constant. Due to the extra degrees of freedom of this theory when compared to GR, this particular choice of the scale factor does not determine uniquely the value of the equation of state. So, for arbitrary $w$, the solutions for $\rho$ and $\psi$ from Eqs.~\eqref{eq:solrho} and \eqref{eq:solpsi} in this particular case become
\begin{equation}\label{eq:solrho-dS}
  \rho = \rho_0  e^{-3\sqrt{\Lambda}\left(t-t_0\right)\left(1+w\right)},
\end{equation}
\begin{equation}\label{eq:solpsi-dS}
  \psi=\psi_0  e^{\frac{3}{2}\sqrt{\Lambda}\left(t-t_0\right)\left(1-w\right)},
\end{equation}
respectively. These solutions are plotted in Figs.~\ref{fig:Plot_dS_Arho} and \ref{fig:Plot_dS_Apsi} for a range of values of the equation of state parameter.


Even though analytic solutions of $\varphi(t)$ and $V_1(\varphi)$ are unattainable without specifying neither $k$ nor $w$, by setting $k=0$ one can find analytic solutions with arbitrary $w$. With this particular choice ($k=0$ and $a(t)$ as given in Eq.~\eqref{eq:a-dS}), Eq.~\eqref{eq:dV1} reduces to a constant, $dV_1/d\varphi=12\Lambda$, which one can integrate  with respect to $\varphi$ and consequently obtain the solution
\begin{equation}\label{eq:solV1-dS-k0}
   V_1(\varphi) = 12\Lambda \varphi,
\end{equation}
where an arbitrary integration constant can be absorbed into $V_0$ in the full expression of the potential, which in this case reads
\begin{equation}\label{eq:solV-dS-k0}
	V(\varphi,\psi) = V_0 + 12\Lambda \varphi+\frac{(1-3w)(1-w)}{(1+3w)}\rho_0 \psi_0^{\frac{2(1+w)}{(1-w)}}\psi^{-\frac{(1+3w)}{(1-w)}}.
\end{equation}	

\begin{figure}
     \includegraphics[width=0.9\columnwidth]{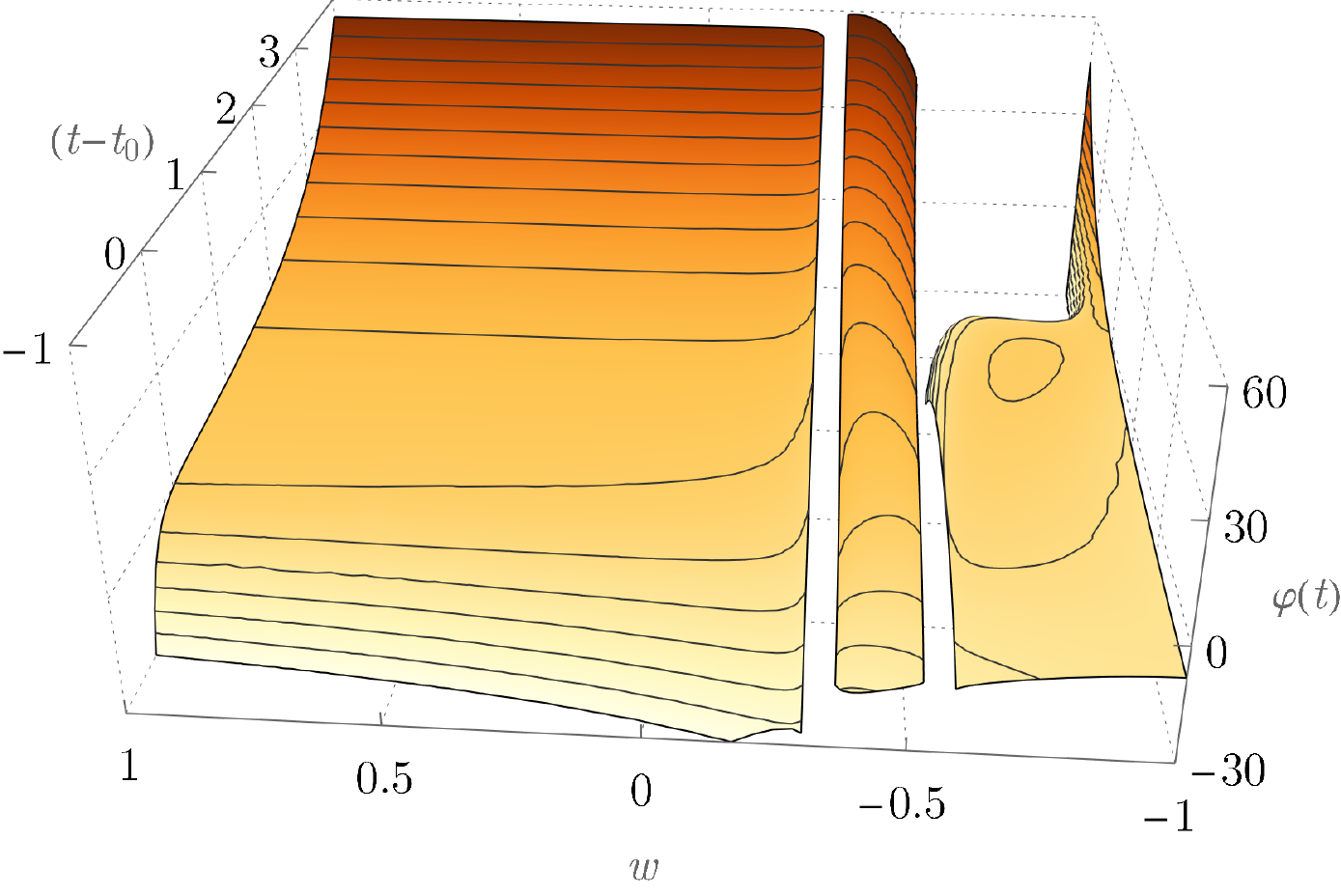}
    \caption{Scalar field $\varphi\left(t\right)$ from Eq.~\eqref{eq:solphi-dSk0} in the particular case where $a\propto e^t$ as given by Eq.~\eqref{eq:a-dS} for a range of values of the equation of state parameter, with $k=V_0=0$ and $\Lambda=\rho_0=\psi_0=\varphi_1=1$.}
    \label{fig:Plot_dS_Aphi_k0}
\end{figure}

Inserting the result of Eq.~\eqref{eq:solV1-dS-k0} into Eq.~\eqref{eq:ttsub} with $k=0$ and solving the resulting equation with respect to $\varphi(t)$ yields
\begin{eqnarray}\label{eq:solphi-dSk0}
		\varphi(t)&=&\varphi_1 e^{\sqrt{\Lambda}\left(t-t_0\right)} -\frac{V_0}{6\Lambda} 
		-\frac{\rho_0}{3\Lambda}\bigg[ \frac{8\pi e^{-3\left(1+w\right)\sqrt{\Lambda}\left(t-t_0\right)}}{\left(4+3w\right)}
		\nonumber \\
	&&	+ \frac{4\left(1+w\right)\psi_0 e^{-\frac{3}{2}\left(1+3w\right)\sqrt{\Lambda}\left(t-t_0\right)}}{\left(1+3w\right)\left(5+9w\right)}\bigg],
\end{eqnarray}
which is undefined at $w=\left\{-4/3,-5/9,-1/3\right\}$, and where $\varphi_1$ is an integration constant. This solution is plotted in Fig.~\ref{fig:Plot_dS_Aphi_k0} for a range of values of $w$.

\begin{figure}
	\includegraphics[width=0.9\columnwidth]{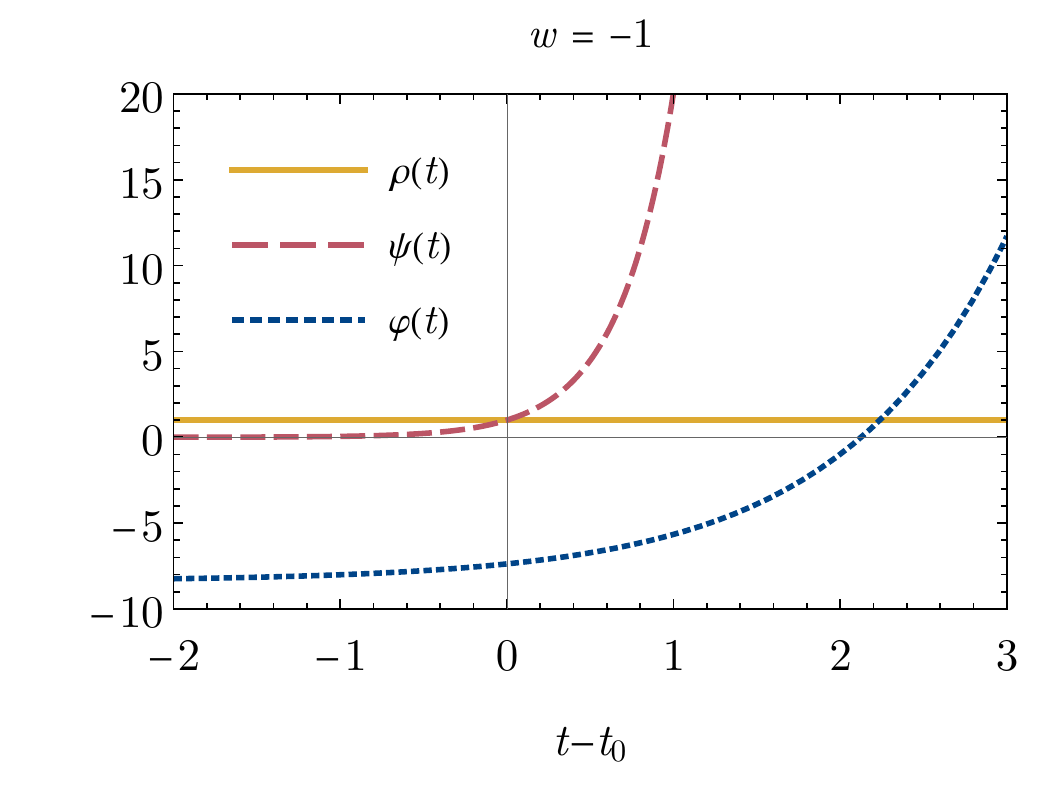}
	\caption{Energy density $\rho(t)$ in Eq.~\eqref{eq:solrho_dS_wneg} and the fields $\psi(t)$ in Eq.~\eqref{eq:solpsi_dS_wneg} and $\varphi(t)$ in Eq.~\eqref{eq:solphi_dS_wneg} when $k=0$ and $w=-1$ in the particular case where $a\propto e^t$ as given by Eq.~\eqref{eq:a-dS}, with $\Lambda=\rho_0=\psi_0=\varphi_1=1$ and $V_0=0$.}
	\label{fig:Plot_dS_Arhophipsi_k0_wneg}
\end{figure}

By way of example, we choose a particular value of the equation of state. Given that in GR a fluid with $w=-1$, which corresponds to a cosmological constant, is associated with an exponential scale factor, this is the value we use here. In the particular case of $w=-1$, Eqs.~\eqref{eq:solrho-dS}, \eqref{eq:solpsi-dS}, \eqref{eq:solV-dS-k0} and \eqref{eq:solphi-dSk0} can be written explicitly as
\begin{equation}\label{eq:solrho_dS_wneg}
	\rho(t)=\rho_0,
\end{equation}
\begin{equation}\label{eq:solpsi_dS_wneg}
	\psi(t)=\psi_0  e^{3\sqrt{\Lambda}\left(t-t_0\right)},
\end{equation}
\begin{equation}
V(\varphi,\psi)= V_0+12\Lambda\varphi-4\rho_0\psi,
\end{equation}
\begin{equation}\label{eq:solphi_dS_wneg}
	\varphi(t)=\varphi_1 e^{\sqrt{\Lambda}\left(t-t_0\right)} -\frac{V_0}{6\Lambda} 
	-\frac{8\pi\rho_0}{3\Lambda},
\end{equation}
respectively. Equations~\eqref{eq:solrho_dS_wneg}, \eqref{eq:solpsi_dS_wneg}, and \eqref{eq:solphi_dS_wneg} are plotted in Fig.~\ref{fig:Plot_dS_Arhophipsi_k0_wneg}.

\begin{figure}
    \includegraphics[width=0.9\columnwidth]{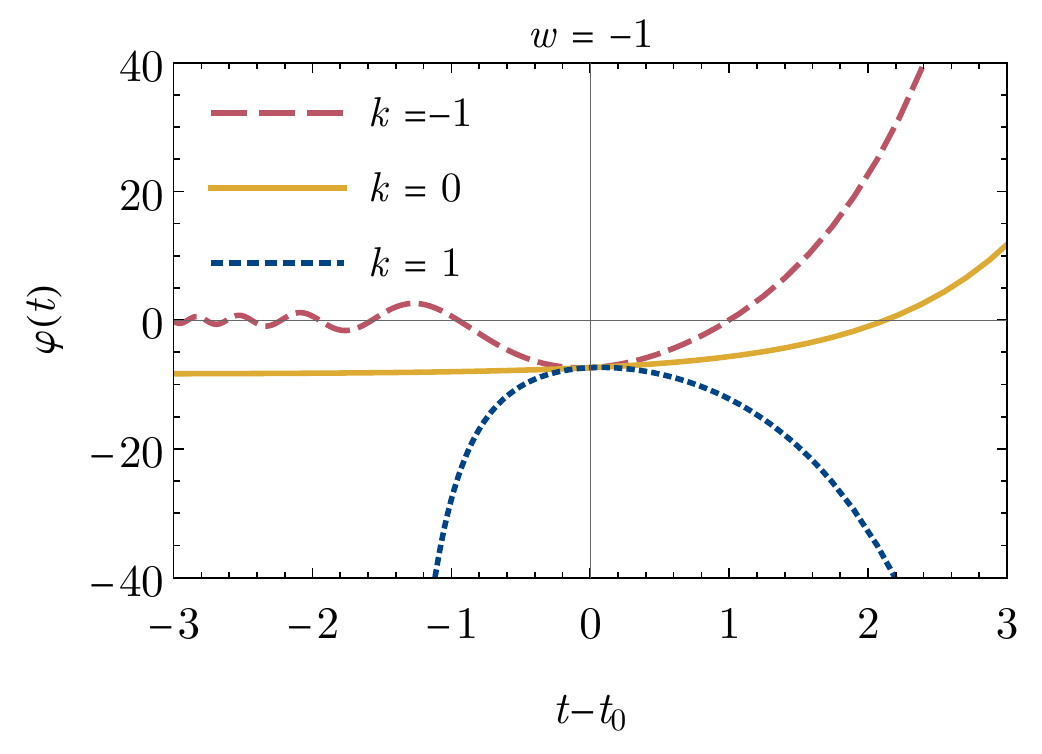}
    \caption{Numerical solutions $\varphi\left(t\right)$ obtained from Eq.~\eqref{eq:dttsub} in the particular case where $a\propto e^t$ as given by Eq.~\eqref{eq:a-dS} the equation of state parameter is $w=-1$, for different values of $k=\{-1,0,1\}$, with $t_0=V_0=0$ and $\Lambda=a_0=\rho_0=\psi_0=\varphi_1=1$. The initial conditions applied were $\varphi(0)=1-8\pi/3$ and $\dot{\varphi}(0)=1$, obtained from Eq.~\eqref{eq:solphi-dSk0}.}
    \label{fig:Plot_dS_Nphi_column}
\end{figure}

So far, we have considered the case when $k=0$. The solutions of $\rho(t)$ and $\psi(t)$ obtained were independent of $k$, but not the solutions of $V_1(\varphi)$ and $\varphi(t)$. Instead, the solutions for $\varphi\left(t\right)$ in non-flat curvature, i.e. for $k=\pm 1$, must be obtained by solving Eq.~\eqref{eq:dttsub} numerically. To this effect, we set $t_0=0$ and $\Lambda=a_0=\rho_0=\psi_0=1$, and we need to further provide two initial conditions. We choose to impose conditions for $\varphi\left(t=t_0\right)$ and $\dot{\varphi}(t=t_0)$ and we inform our choice of values from the analytic solution we already have in Eq.~\eqref{eq:solphi-dSk0}, by also setting $\varphi_1=1$ and $V_0=0$. Thus, the initial conditions are such that the solutions for $k=\pm 1$ are consistent with the ones for $k=0$ at $t=t_0=0$, which is useful for comparison of the results. In this way, for the $w=-1$ case, we used $\varphi(0)=1-8\pi/3$ and $\dot{\varphi}(0)=1$ as conditions to solve Eq.~\eqref{eq:dttsub} numerically. The numerical solutions obtained are plotted in Fig.~\ref{fig:Plot_dS_Nphi_column}.

\subsection{Power-law expansion with $a\propto t^{1/2}$}\label{subsec:rad}

\begin{figure}
	\includegraphics[width=0.9\columnwidth]{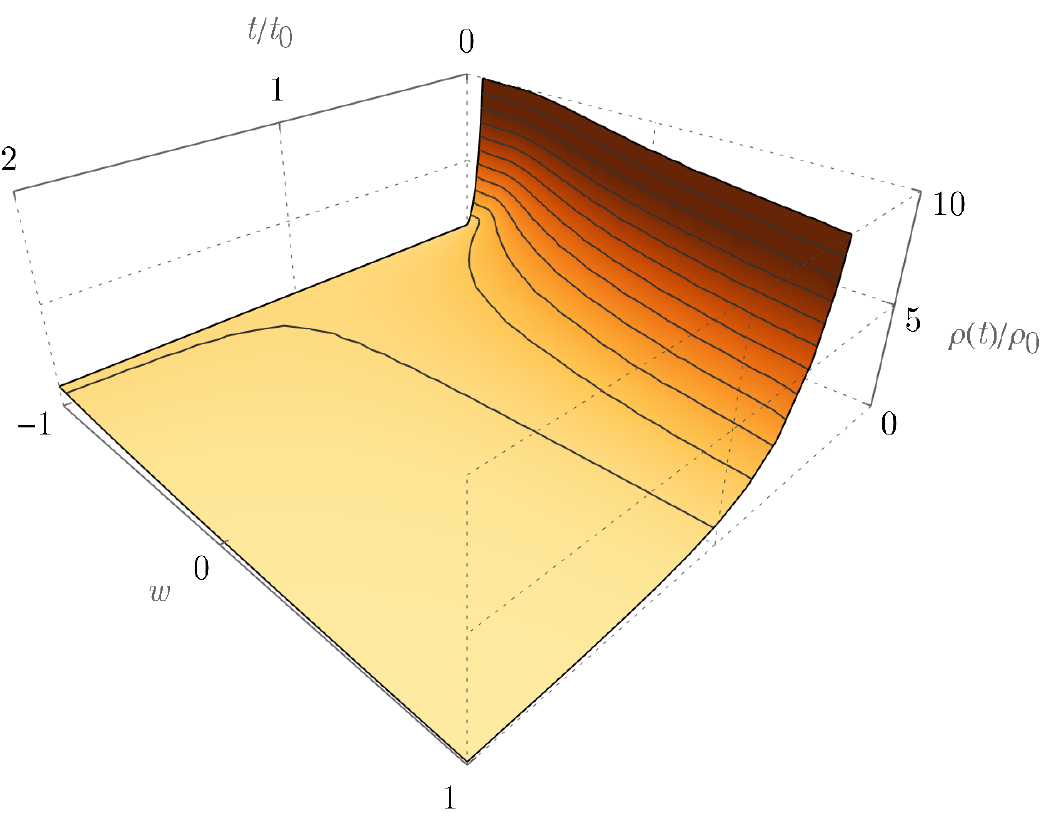}
	\caption{Energy density $\rho\left(t\right)$ from Eq.~\eqref{eq:solrho-r} in the particular case where $a \propto t^{1/2}$ as given by Eq.~\eqref{eq:a-r} for a range of values of the equation of state parameter $-1<w<1$.}
	\label{fig:Plot3D_rho_r}
\end{figure}

\begin{figure}
	\includegraphics[width=0.9\columnwidth]{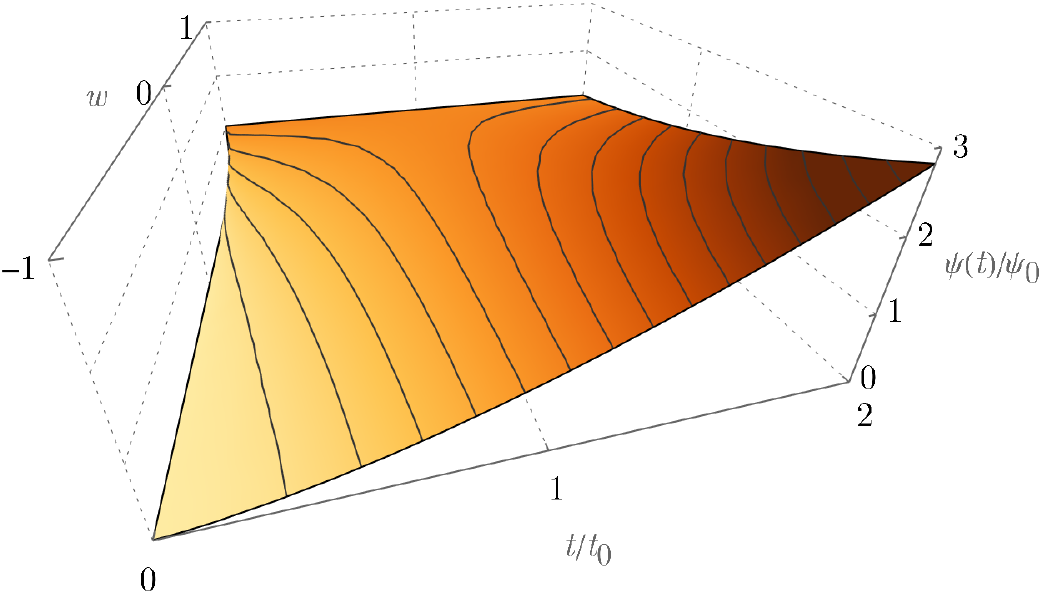}
	\caption{Scalar field $\psi\left(t\right)$ from Eq.~\eqref{eq:solpsi-r} in the particular case where  $a \propto t^{1/2}$ as given by Eq.~\eqref{eq:a-r} for a range of values of the equation of state parameter.}
	\label{fig:Plot3D_psi_r}
\end{figure}

In this section, we consider a scale factor that increases as a power-law motivated by the GR behavior of radiation dominated universes, viz. a scale factor of the form
\begin{equation}\label{eq:a-r}
    a(t)=a_0 \left(\frac{t}{t_0}\right)^{\frac{1}{2}},
\end{equation}
where $a_0$ and $t_0$ are constants defined in such a way that $a_0$ is the value of the scale factor when $t=t_0$, which can be taken to be the present time, for instance. With this particular choice, the solutions for $\rho$ and $\psi$ from Eqs.~\eqref{eq:solrho} and \eqref{eq:solpsi} become
\begin{equation}\label{eq:solrho-r}
    \rho = \rho_0 \left(\frac{t}{t_0}\right)^{-\frac{3}{2}\left(1+w\right)} ,
\end{equation}
\begin{equation}\label{eq:solpsi-r}
   \psi=\psi_0 \left(\frac{t}{t_0}\right)^{\frac{3}{4}\left(1-w\right)} ,
\end{equation}
respectively. These solutions are plotted for a range of values of the equation of state in Figs.~\ref{fig:Plot3D_rho_r} and \ref{fig:Plot3D_psi_r} for different equations of state. 


\begin{figure}
    \includegraphics[width=0.9\columnwidth]{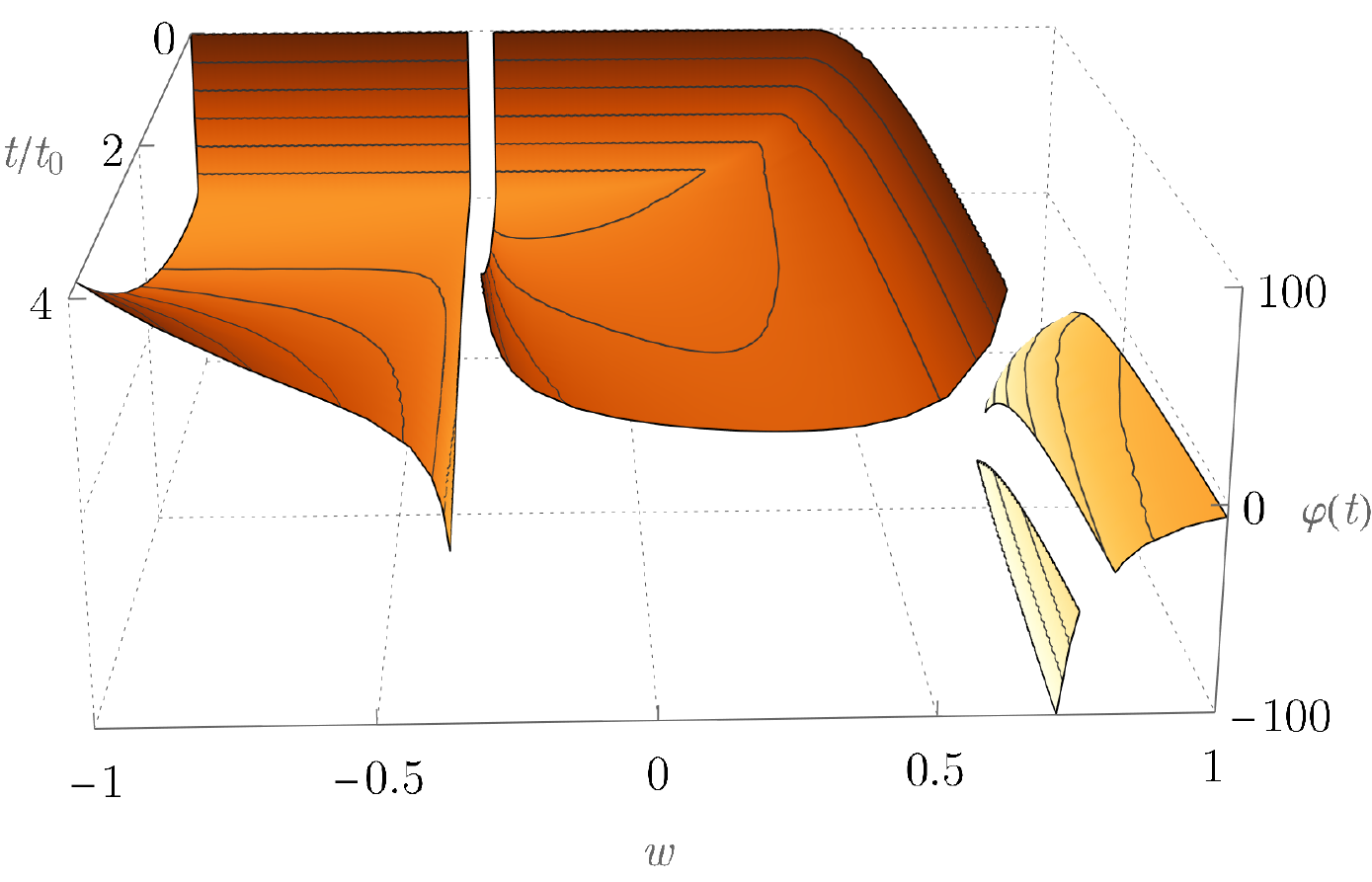}
    \caption{Scalar field $\varphi\left(t\right)$ from Eq.~\eqref{eq:solphi-rk0} in the particular case where  $a \propto t^{1/2}$ as given by Eq.~\eqref{eq:a-r} for a range of the equation of state parameter, with $t_0=\rho_0=\psi_0=\varphi_1=V_0=1$.}
    \label{fig:Plot3D_varphi_r}
\end{figure}

We now follow a similar procedure as in the previous Sec.~\ref{subsec:dS}. With the particular form of the scale factor of Eq.~\eqref{eq:a-r}, the right-hand side of Eq.~\eqref{eq:dV1} vanishes for $k=0$, which means there is no $\varphi$ dependence in $V(\varphi,\psi)$, i.e. $V_1(\varphi) = 0$, and so,
\begin{equation}\label{eq:solV-r-k0}
V(\varphi,\psi) = V_0 +\frac{(1-3w)(1-w)}{(1+3w)}\rho_0 \psi_0^{\frac{2(1+w)}{(1-w)}}\psi^{-\frac{(1+3w)}{(1-w)}}.
\end{equation}	
Substituting $V_1(\varphi) = 0$ into Eq.~\eqref{eq:ttsub} and solving with respect to $\varphi(t)$ leads to
\begin{eqnarray}\label{eq:solphi-rk0}
   \varphi(t) &=& \varphi_1\left(\frac{t}{t_0}\right)^{-\frac{1}{2}} + \frac{2V_0 t_0^2}{15}\left(\frac{t}{t_0}\right)^2  
   \nonumber \\
  && +\frac{16\rho_0 t_0^2}{3}  \Bigg[ \frac{2\pi}{\left(2-3w\right)} \left(\frac{t}{t_0}\right)^{\frac{\left(1-3w\right)}{2}}  
   \nonumber \\
    && +   \frac{\left(1+w\right)\psi_0}{\left(1+3w\right)\left(7-9w\right)} \left(\frac{t}{t_0}\right)^{\frac{\left(5-9w\right)}{4}}  \Bigg],
\end{eqnarray}
where $\varphi_1$ is an integration constant. This solution, which is undefined at $w=\left\{-1/3,2/3,7/9\right\}$, is plotted in Fig.~\ref{fig:Plot3D_varphi_r}. 

\begin{figure}
	\includegraphics[width=0.9\columnwidth]{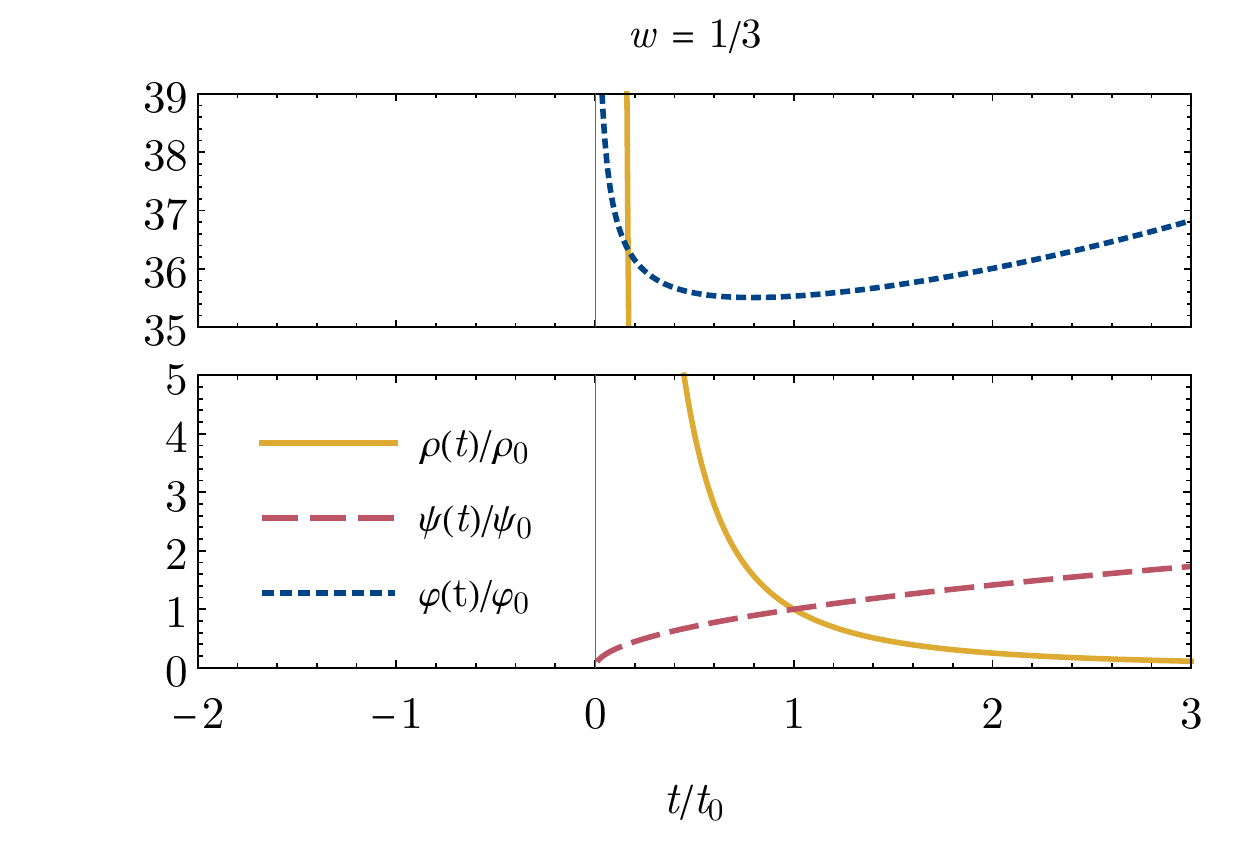}
	\caption{Energy density $\rho(t)$ in Eq.~\eqref{eq:solrho_r_wpos} and the fields $\psi(t)$ in Eq.~\eqref{eq:solpsi_r_wpos} and $\varphi(t)$ in Eq.~\eqref{eq:solphi_r_wpos} when $k=0$ and $w=1/3$ in the particular case where  $a \propto t^{1/2}$ as given by Eq.~\eqref{eq:a-r}, with $\rho_0=\psi_0=\varphi_1=V_0=1$.}
	\label{fig:Plot_r_Arhophipsi_k0_wpos}
\end{figure}

Again, we choose a particular value of the equation of state, as an example. Given that in GR a fluid with the scale factor evolution as given by Eq.~\eqref{eq:a-r} is associated with a radiation era, here we choose $w=1/3$. In this particular case, Eqs.~\eqref{eq:solrho-r}, \eqref{eq:solpsi-r},  \eqref{eq:solV-r-k0} and \eqref{eq:solphi-rk0} can be written explicitly as
\begin{equation}\label{eq:solrho_r_wpos}
 \rho(t) = \rho_0 \left(\frac{t}{t_0}\right)^{-2},
\end{equation}
\begin{equation}\label{eq:solpsi_r_wpos}
\psi(t)=\psi_0 \left(\frac{t}{t_0}\right)^{\frac{1}{2}},
\end{equation}
\begin{equation}
V(\varphi,\psi) = V_0,
\end{equation}
\begin{equation}\label{eq:solphi_r_wpos}
\begin{multlined}
\varphi(t)=\varphi_1\left(\frac{t}{t_0}\right)^{-\frac{1}{2}} + \frac{2V_0 t_0^2}{15}\left(\frac{t}{t_0}\right)^2 \\
+\frac{16\rho_0 t_0^2}{3}  \left[  2\pi 
+   \frac{\psi_0}{6} \left(\frac{t}{t_0}\right)^{\frac{1}{2}}  \right],
\end{multlined}
\end{equation}
respectively. Equations~\eqref{eq:solrho_r_wpos}, \eqref{eq:solpsi_r_wpos} and \eqref{eq:solphi_r_wpos} are plotted in Fig.~\ref{fig:Plot_r_Arhophipsi_k0_wpos}.

\begin{figure}
    \includegraphics[width=0.9\columnwidth]{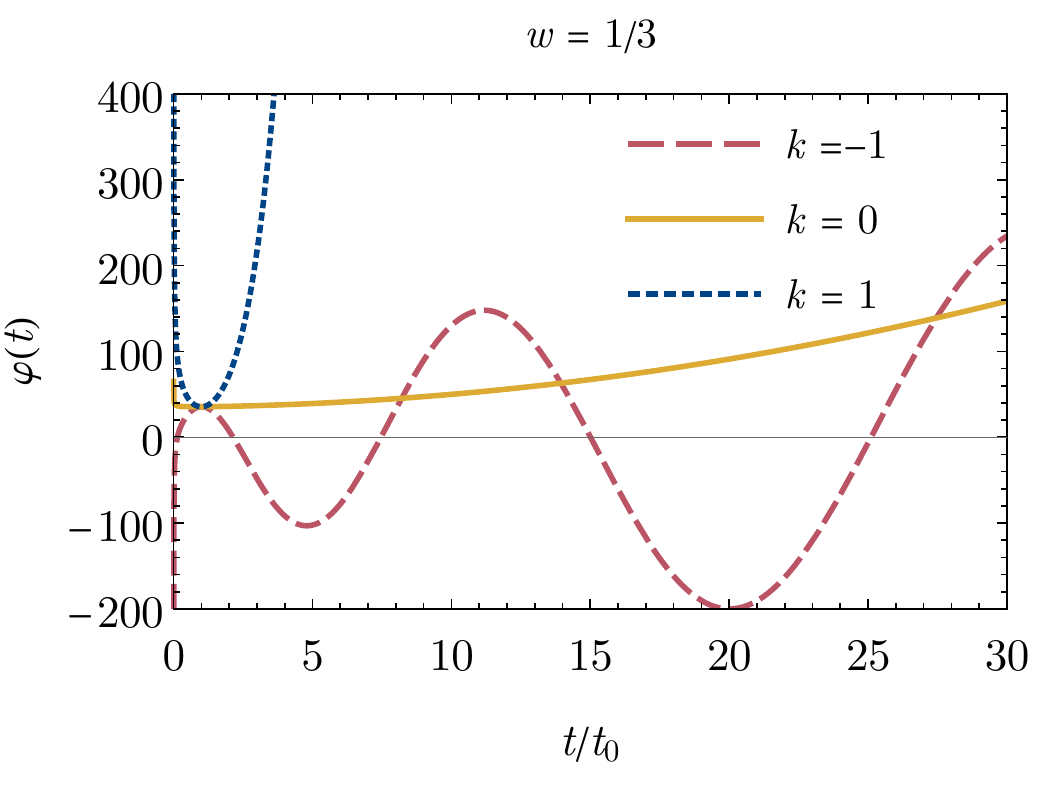}    
    \caption{Numerical solutions of $\varphi\left(t\right)$, in the particular case where  $a \propto t^{1/2}$ as given by Eq.~\eqref{eq:a-r}, obtained by solving Eq.~\eqref{eq:dttsub} with the initial conditions ${\varphi\left(t_0\right)=91/45+32\pi/3}$ and ${\dot{\varphi}\left(t_0\right)=19/90}$, for $w=1/3$, and for different values of $k=\{-1,0,1\}$, with $t_0=a_0=\rho_0=\psi_0=\varphi_1=V_0=1$.}
    \label{fig:Plot_r_Nphi_column}
\end{figure}

Following the approach from the previous subsection, we find numerical solutions for non-flat geometry, ${k=\pm 1}$, using initial conditions obtained in such a way to guarantee that these solutions are consistent with the analytical ones obtained for $k=0$ at $t=t_0$. For the $w=1/3$ case, setting $\rho_0=\psi_0=\varphi_1=V_0=1$ in Eq.~\eqref{eq:solphi_r_wpos} and its derivative, the initial conditions we use are $\varphi\left(t_0\right)=91/45+32\pi/3$ and $\dot{\varphi}\left(t_0\right)=19/90$. The numerical solutions obtained are plotted in Fig.~\ref{fig:Plot_r_Nphi_column}. 

\subsection{Power-law expansion with $a\propto t^{2/3}$}\label{subsec:mat}

Finally, in this section we consider a scale factor increasing as a power-law motivated by the behavior of matter-dominated universes in GR, i.e. we consider the scale factor in the form
\begin{equation}\label{eq:a-m}
    a(t)=a_0 \left(\frac{t}{t_0}\right)^{\frac{2}{3}},
\end{equation}
where $a_0$ and $t_0$ are constants chosen in such a way that $a_0$ represents the scale factor at the time $t_0$, which can be taken to represent the present cosmological time. With this particular choice, the solutions for $\rho$ and $\psi$ from Eqs.~\eqref{eq:solrho} and \eqref{eq:solpsi} become
\begin{equation}\label{eq:solrho-m}
    \rho(t) = \rho_0 \left(\frac{t}{t_0}\right)^{-2(1+w)},
\end{equation}
\begin{equation}\label{eq:solpsi-m}
   \psi(t)=\psi_0\left(\frac{t}{t_0}\right)^{\left(1-w\right)},
\end{equation}
respectively. These functions are not plotted here as they are qualitatively similar to those in Figs.~\ref{fig:Plot3D_rho_r} and \ref{fig:Plot3D_psi_r}.

\begin{figure}
    \includegraphics[width=0.9\columnwidth]{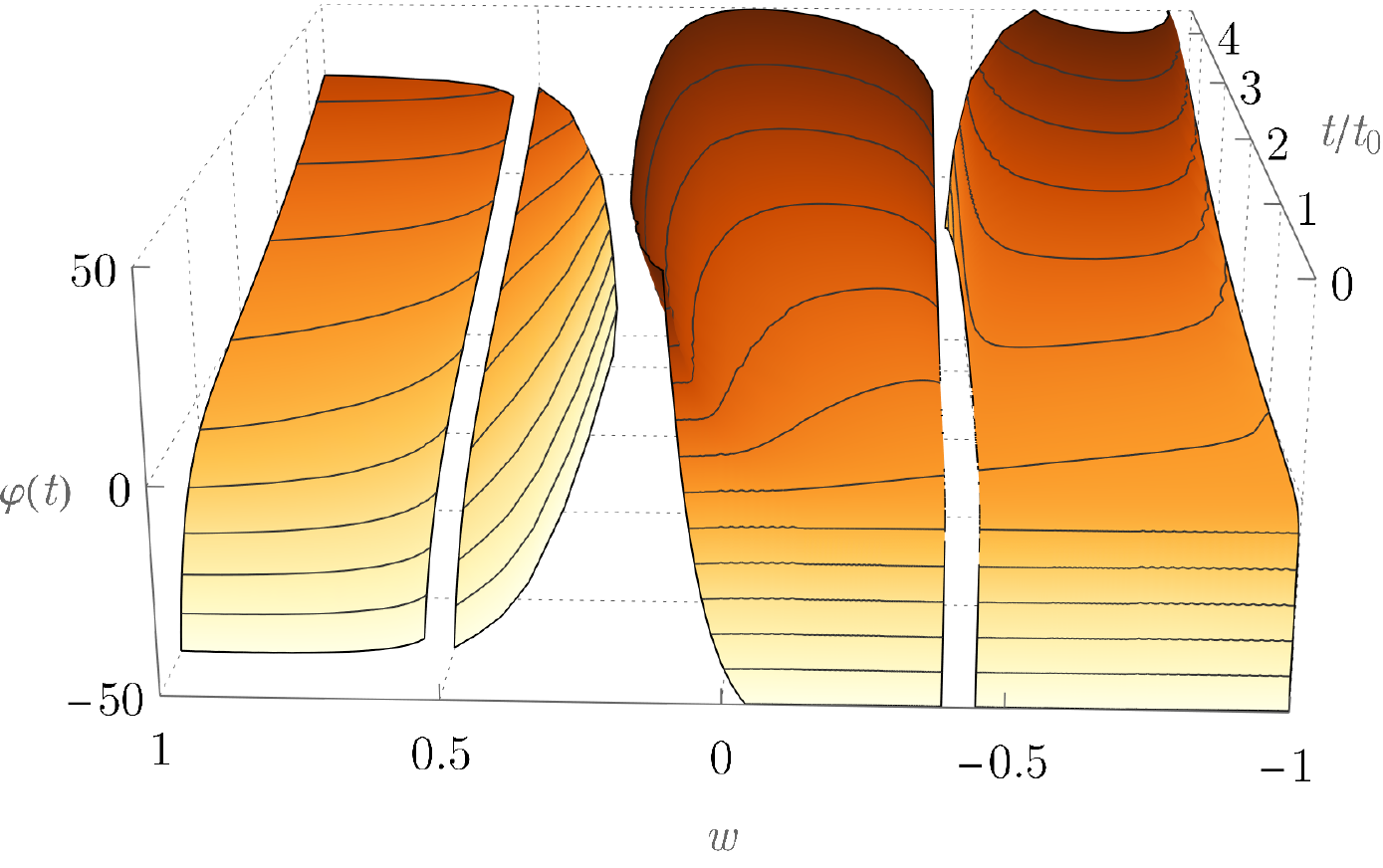}
    \caption{Scalar field $\varphi\left(t\right)$ from Eq.~\eqref{eq:solphi-m-k0} in the particular case where ${a \propto t^{2/3}}$ as given by Eq.~\eqref{eq:a-m} for a range of values of the equation of state parameter, setting $t_0=\rho_0=\psi_0=\varphi_1=1$ and $\varphi_2=-1$.}
    \label{fig:Plot_m_Aphi_k0}
\end{figure}

In the next steps, a slightly different method needs to be taken in this particular case. When $a(t)$ is of the form given in Eq.~\eqref{eq:a-m} and $k=0$, Eq.~\eqref{eq:dV1}  becomes 
\begin{equation}\label{eq:dV1-mk0}
    \frac{d V_1}{d\varphi}=\frac{4}{3t^2}.
\end{equation}
This explicit dependence on $t$ prevents one to integrate this equation and to obtain the solution for $V_1\left(\varphi\right)$ at this stage, as was done for the other cases. One must first solve Eq.~\eqref{eq:dttsub} to obtain the explicit solution of $\varphi\left(t\right)$, and only if this solution is invertible can one obtain $V_1\left(\varphi\right)$. When $k=0$, Eq.~\eqref{eq:dttsub} yields the solution
\begin{eqnarray}\label{eq:solphi-m-k0}
        \varphi(t)&=&\varphi_1\left(\frac{t}{t_0}\right)^{\frac{5+\sqrt{73}}{6}}+\varphi_2\left(\frac{t}{t_0}\right)^{\frac{5-\sqrt{73}}{6}}
        \nonumber \\
        && +\frac{12\pi\left(1+w\right)\rho_0 t_0^2}{\left[2-w\left(5+6w\right)\right]}\left(\frac{t}{t_0}\right)^{-2w}
        \nonumber \\
        && +\frac{\left(1+w\right)\rho_0\psi_0 t_0^2}{\left[2+w\left(1-9w\right)\right]}\left(\frac{t}{t_0}\right)^{\left(1-3w\right)},  
\end{eqnarray}
where $\varphi_1$ and $\varphi_2$ are integration constants. This function is undefined at ${w=\left\{-\left(5\pm\sqrt{73}\right)/12, \left(1\pm\sqrt{73}\right)/18\right\}}$. This solution is plotted in Fig.~\ref{fig:Plot_m_Aphi_k0} for a range of values of the equation of state, choosing opposite signs for the integration constants, ${\varphi_1=+1}$ and ${\varphi_2=-1}$ so that the function is injective in most of the domain under consideration.

\begin{figure}
	\includegraphics[width=0.9\columnwidth]{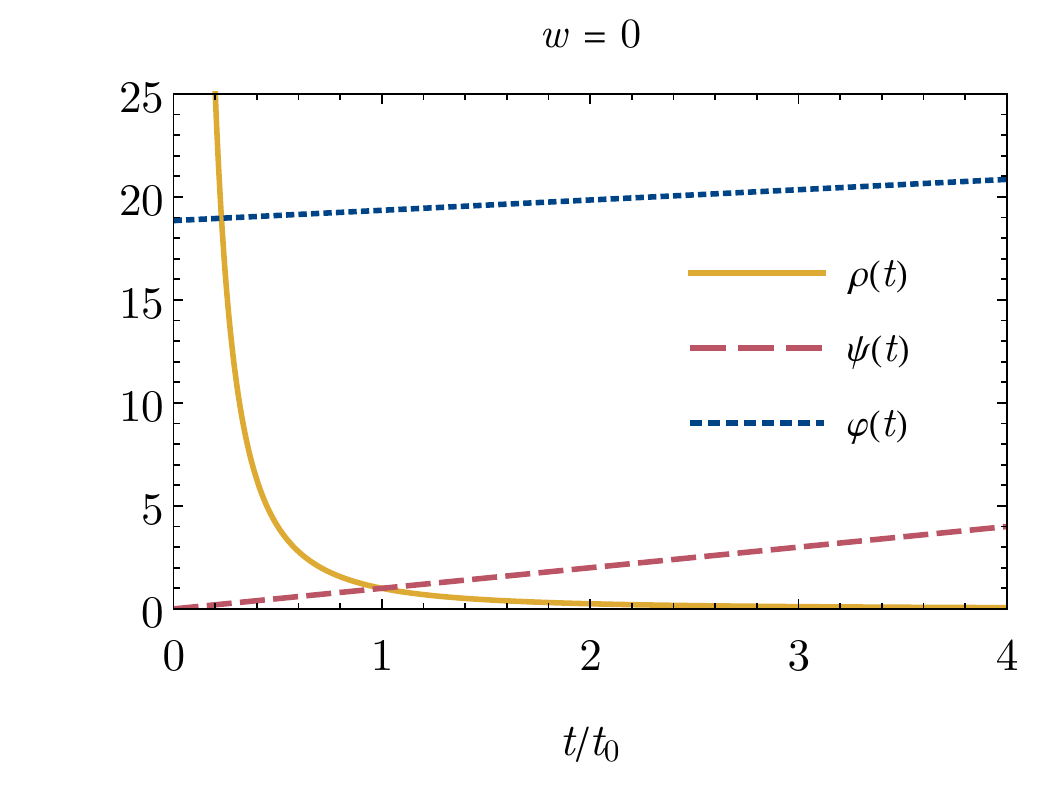}
	\caption{Energy density $\rho(t)$ in Eq.~\eqref{eq:solrho_m_w0} and the fields $\psi(t)$ in Eq.~\eqref{eq:solpsi_m_w0} and $\varphi(t)$ in Eq.~\eqref{eq:solphi_m_k0_w0} when ${k=w=\varphi_1=\varphi_2=0}$ in the particular case where  $a \propto t^{2/3}$ as given by Eq.~\eqref{eq:a-m}, with $t_0=\rho_0=\psi_0=1$.}
	\label{fig:Plot_m_Arhophipsi_k0_w0}
\end{figure}

\begin{figure}
    \includegraphics[width=0.8\columnwidth]{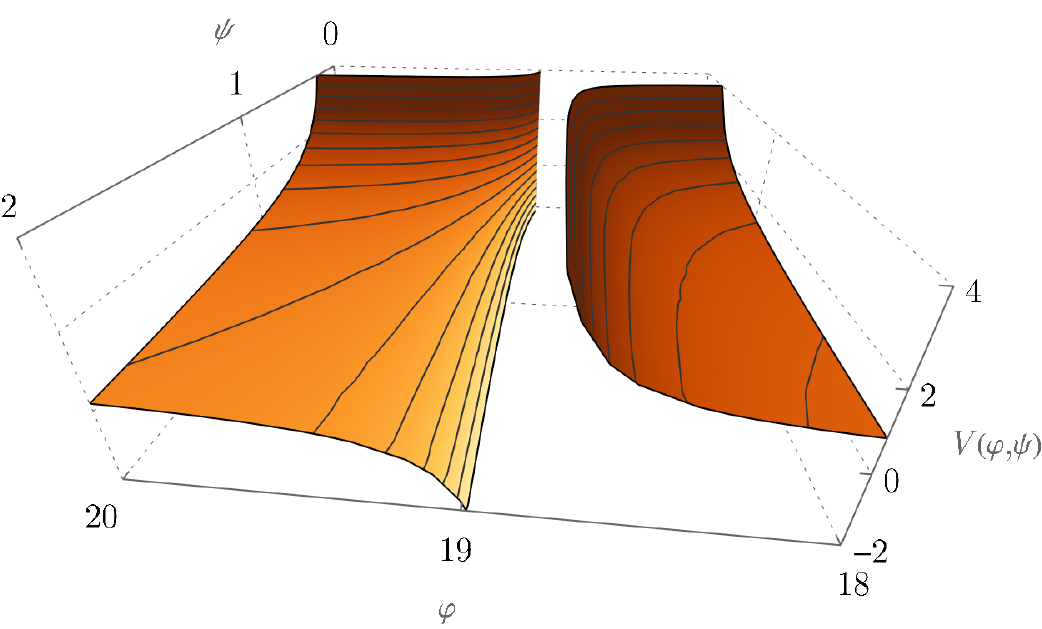}
    \caption{Potential $V\left(\varphi,\psi\right)$ from Eq.~\eqref{eq:solV-mk0w0-phi1phi20} in the particular case where $a \propto t^{2/3}$ as given by Eq.~\eqref{eq:a-m} and ${k=w=\varphi_1=\varphi_2=0}$, setting $t_0=\rho_0=\psi_0=1$ and $V_0=0$.}
    \label{fig:Plot3D_V_m}
\end{figure}

Having obtained the solutions for $\varphi\left(t\right)$, the form of the potential function $V_1\left(\varphi\right)$ can be obtained if the function $\varphi\left(t\right)$ is invertible to get $t(\varphi)$ in order to integrate Eq.~\eqref{eq:dV1-mk0}. By way of example, let us analyze this situation for the case of pressureless (or dust-like) matter, i.e. ${w=0}$, which in GR is the equation of state associated with the scale factor evolution as given by Eq.~\eqref{eq:a-m}. If $w=0$ in Eq.~\eqref{eq:solphi-m-k0}, we can proceed analytically if we choose $\varphi_1=\varphi_2=0$, so that there remains only one $t$-dependent term, and we can invert the equation. With this choice of constants, Eqs.~\eqref{eq:solrho-m}, \eqref{eq:solpsi-m} and \eqref{eq:solphi-m-k0} give, respectively,
\begin{equation}\label{eq:solrho_m_w0}
\rho(t) = \rho_0 \left(\frac{t}{t_0}\right)^{-2},
\end{equation}
\begin{equation}\label{eq:solpsi_m_w0}
\psi(t)=\psi_0\left(\frac{t}{t_0}\right),
\end{equation}
\begin{equation}\label{eq:solphi_m_k0_w0}
\varphi(t)=
6\pi\rho_0 t_0^2+\frac{\rho_0 \psi_0 t_0^2}{2}\left(\frac{t}{t_0}\right),  
\end{equation}
These particular solutions are plotted in Fig.~\ref{fig:Plot_m_Arhophipsi_k0_w0}.
Inverting Eq.~\eqref{eq:solphi_m_k0_w0} to find $t\left(\varphi\right)$ to substitute into Eq.~\eqref{eq:dV1-mk0}, and integrating with respect to $\varphi$, we obtain the potential
\begin{equation}\label{eq:solV-mk0w0-phi1phi20}
    V(\varphi,\psi)=V_0+\frac{\left(\rho_0\psi_0 t_0\right)^2}{18\pi\rho_0t_0^2-3\varphi}+\frac{\rho_0\psi_0^2}{\psi},
\end{equation}
which is undefined at ${\varphi=6\pi\rho_0 t_0^2}$ and at ${\psi=0}$. This form of the potential is shown in Fig.~\ref{fig:Plot3D_V_m}, in the region near the discontinuities in $\varphi$ and $\psi$.

\begin{figure}
    \includegraphics[width=0.9\columnwidth]{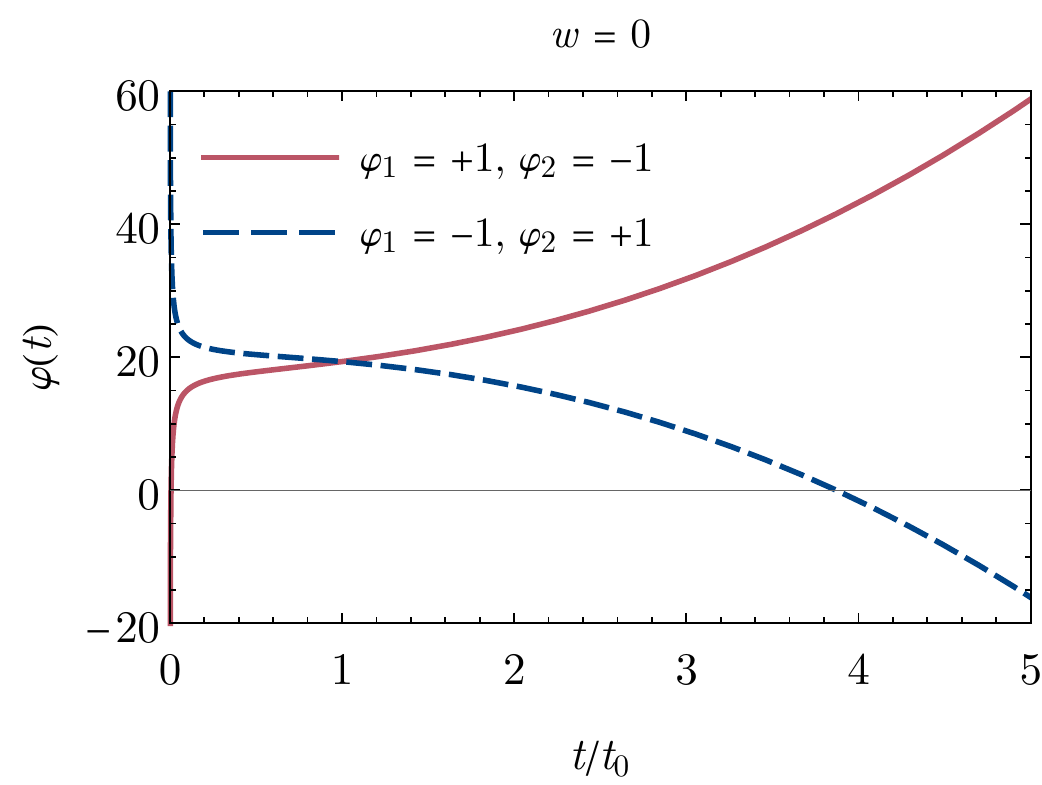}
    \caption{Scalar field $\varphi\left(t\right)$ from Eq.~\eqref{eq:solphi-m-k0} in the particular case where $a \propto t^{2/3}$ as given by Eq.~\eqref{eq:a-m} for two different combinations of $\varphi_1$ and $\varphi_2$ with $t_0=\rho_0=\psi_0=1$ and $w=0$.}
    \label{fig:Plot_m_Nphi_k0w0}
\end{figure}

\begin{figure}
    \includegraphics[width=0.9\columnwidth]{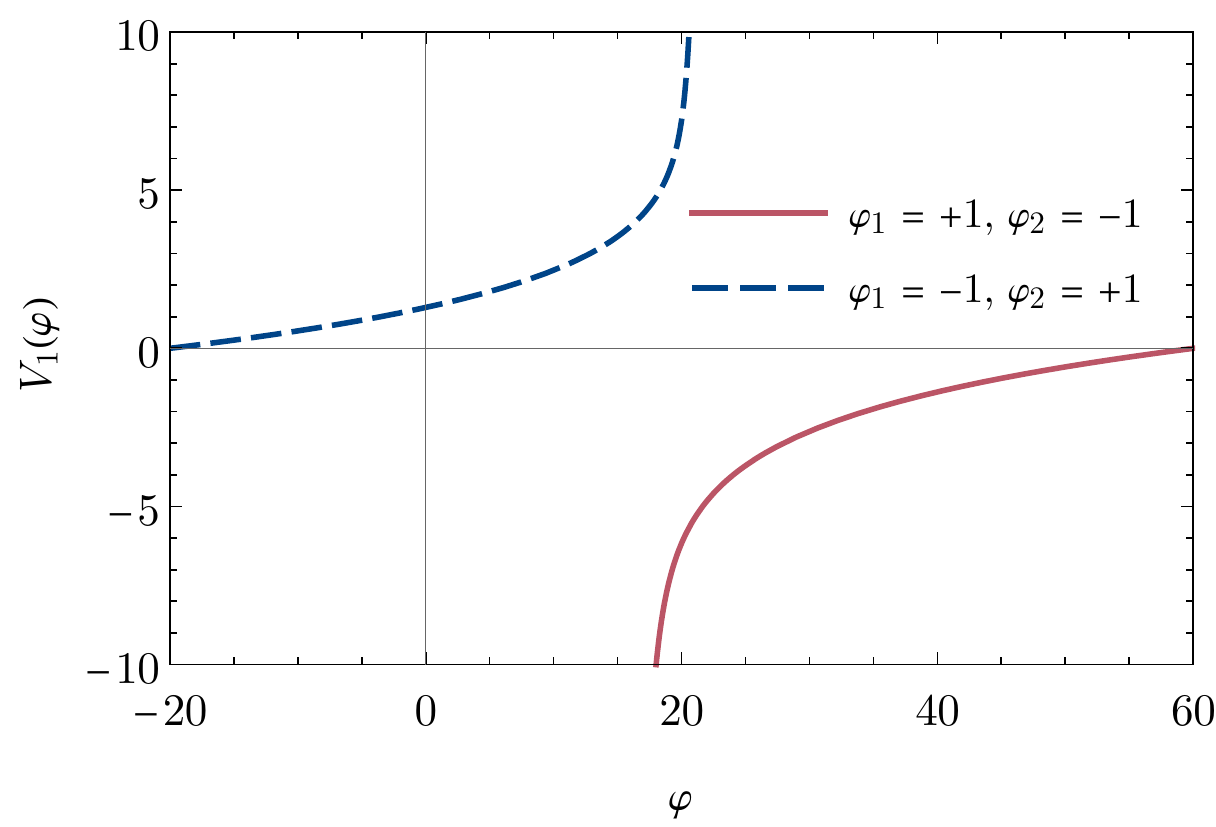}
    \caption{Numerical solutions of  $V_1(\varphi)$ with an arbitrary initial condition $V_1(60)=0$ when $\varphi_1=+1,\varphi_2=-1$ and $V_1(-20)=0$ when $\varphi_1=-1,\varphi_2=+1$ (chosen for visualization purposes), in the particular case where $a\propto t^{2/3}$ as given by Eq.~\eqref{eq:a-m} with $w=0$ and $t_0=\rho_0=\psi_0=1$.}
    \label{fig:Plot_m_NV1_k0w0}
\end{figure}

\begin{figure}
	\includegraphics[width=0.8\columnwidth]{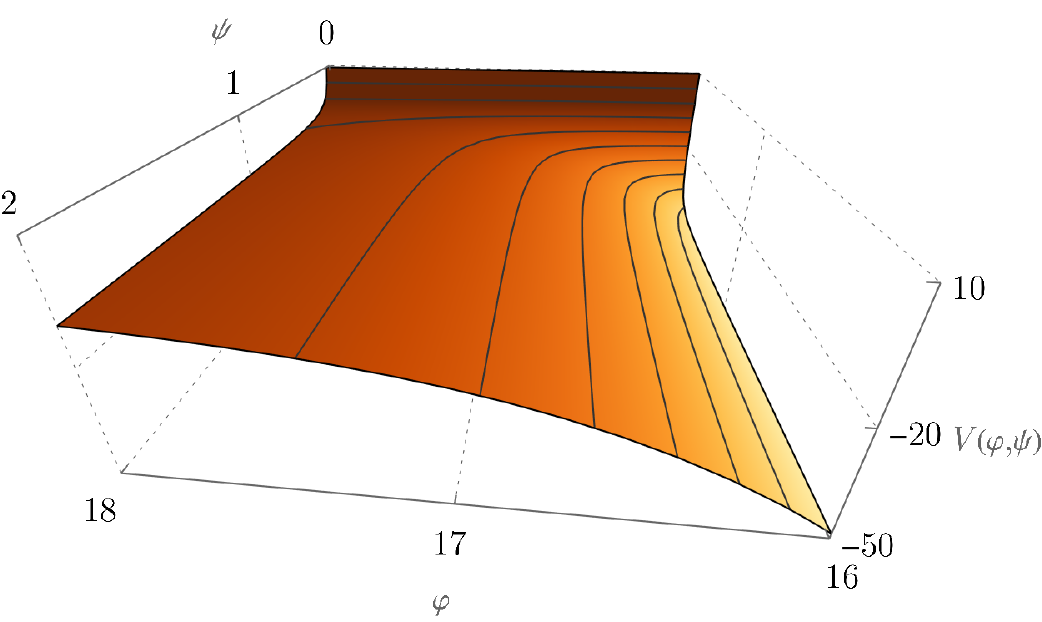}
	\includegraphics[width=0.8\columnwidth]{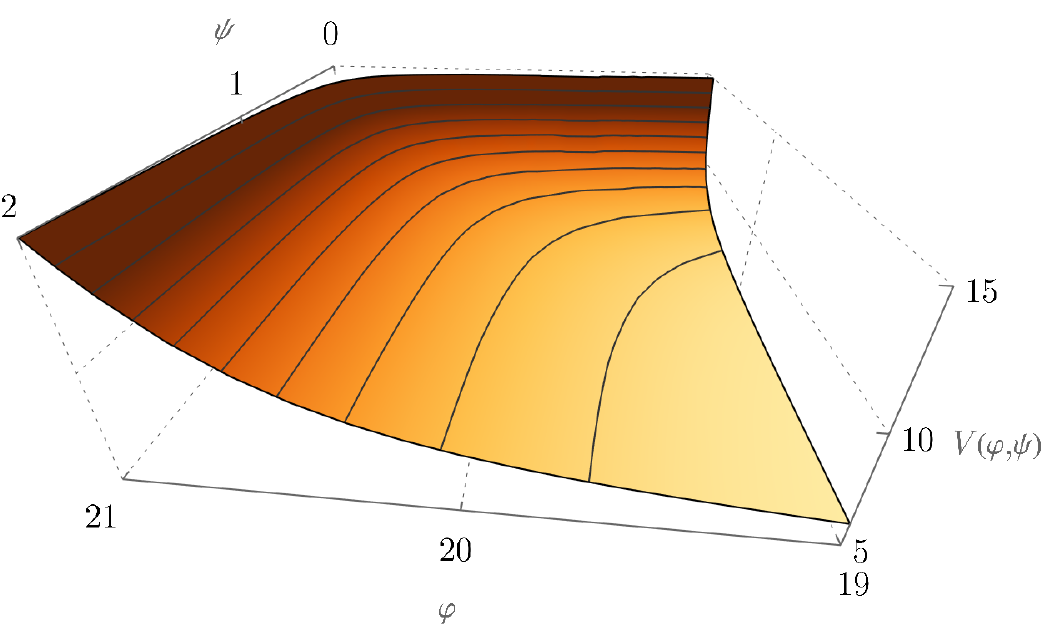}
	\caption{Potential $V\left(\varphi,\psi\right)=V_0+V_1(\varphi)+V_2(\psi)$ in the particular case when $a\propto t^{2/3}$ as given by Eq.~\eqref{eq:a-m} and ${k=w=0}$, setting ${V_0=0}$, $V_1(\varphi)$ as the numerical solutions in Fig.~\ref{fig:Plot_m_NV1_k0w0} with ${\varphi_1=+1 \land \varphi_2=-1}$ (top panel) and ${\varphi_1=-1 \land \varphi_2=+1}$ (bottom panel), and $V_2(\psi)$ as given by Eq.~\eqref{eq:solV2_psi} with $\rho_0=\psi_0=1$.}
	\label{fig:Plot3D_V_m_phi12}
\end{figure}

This is as far as we can go analytically. However, we may like to have solutions for the potential without requiring any term to vanish in the expression of $\varphi(t)$, i.e. without requiring $\varphi_1=\varphi_2=0$ in Eq.~\eqref{eq:solphi-m-k0}.  As long as we guarantee that $\varphi(t)$ is injective, we can calculate the inverse function numerically. Then we will be able to solve for $V_1(\varphi)$, and consequently, find the corresponding potential $V(\varphi,\psi)$ for each case. In the case when $w=0$, $\varphi(t)$ is injective if $\varphi_1$ and $\varphi_2$ have opposite signs, i.e. if either $\varphi_1>0 \land \varphi_2<0 $, or $\varphi_1<0 \land \varphi_2>0$. These two choices are plotted in Fig.~\ref{fig:Plot_m_Nphi_k0w0}. By calculating numerically the inverse functions of these forms of $\varphi(t)$, and using Eq.~\eqref{eq:dV1-mk0}, we find the corresponding numerical solutions of $V_1(\varphi)$ which are shown in Fig.~\ref{fig:Plot_m_NV1_k0w0}. A 3D plot of the corresponding $V(\varphi,\psi)$ is shown in Fig.~\ref{fig:Plot3D_V_m_phi12}. 


\begin{figure}
	\includegraphics[width=0.9\columnwidth]{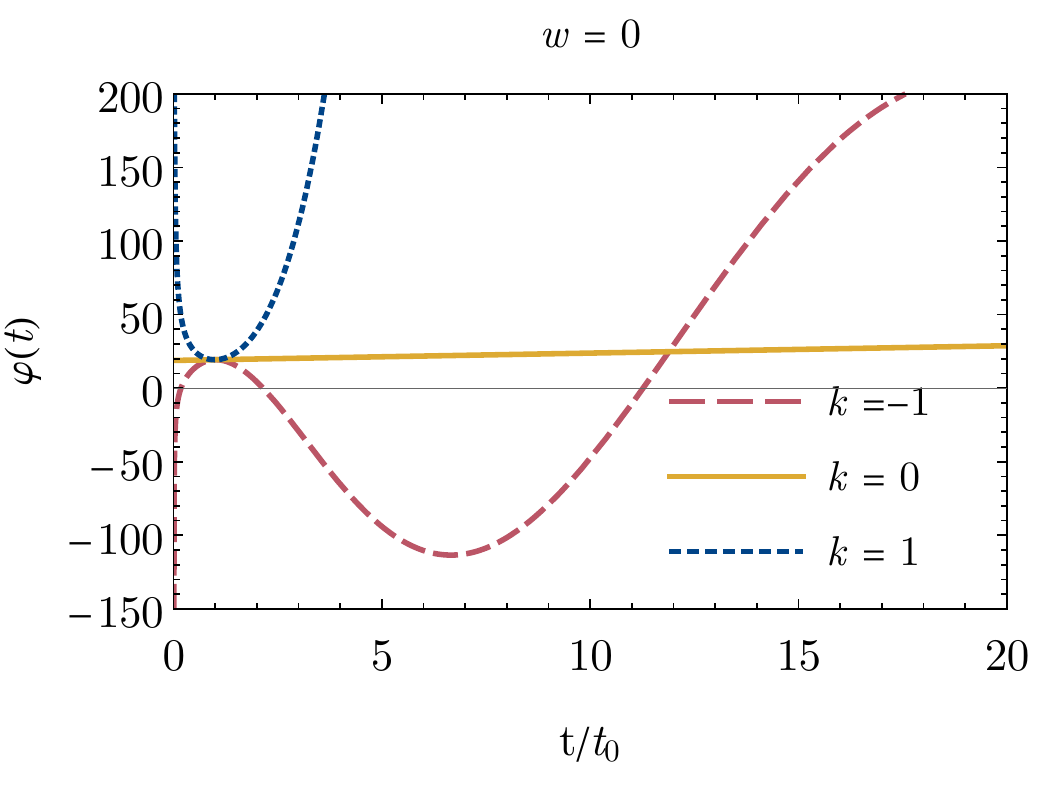}    
	\caption{Numerical solutions of $\varphi\left(t\right)$, in the particular case where  $a \propto t^{2/3}$ as given by Eq.~\eqref{eq:a-m}, obtained by solving Eq.~\eqref{eq:dttsub} with the initial conditions ${\varphi\left(t_0\right)=1/2+6\pi}$ and ${\dot{\varphi}\left(t_0\right)=1/2}$, for $w=0$, and for different values of $k=\{-1,0,1\}$, with $t_0=a_0=\rho_0=\psi_0=1$ and $V_0=0$.}
	\label{fig:Plot_m_Nphi_w0}
\end{figure}

To conclude our analysis of this particular case, as in previous sections, we look for numerical solutions in non-flat geometry. Taking, for instance, Eq.~\eqref{eq:solphi_m_k0_w0} as  the source for the initial conditions, we use ${\varphi\left(t_0\right)=1/2+6\pi}$ and ${\dot{\varphi}\left(t_0\right)=1/2}$ to solve Eq.~\eqref{eq:dttsub} numerically for $k=\pm1$. Figure~\ref{fig:Plot_m_Nphi_w0} shows the results, including the $k=0$ case as given by Eq.~\eqref{eq:solphi_m_k0_w0}.

We have thus presented complete solutions for the system's variables in several particular cases. As a next step, it is interesting to find the explicit form of the function $f\left(R,T\right)$ which is consistent with these cosmologies. 

\section{Explicit form of $f(R,T)$}\label{sec:fRT-explicit}

The forms of the function $f\left(R,T\right)$ associated with the potentials $V\left(\varphi,\psi\right)$ derived throughout this work can be obtained from the definition of the potential given in Eq.~\eqref{eq:potential}. To do so, one sets $\alpha=R$ and $\beta=T$, and uses the definitions $\varphi\equiv f_R$ and $\psi\equiv f_T$. The result is a partial differential equation for $f\left(R,T\right)$ as
\begin{equation}\label{eq:potentialRT}
f(R,T)=-V(f_R,f_T)+f_R R +f_T T.
\end{equation}
Taking the partial derivatives of this equation with respect to $R$ and $T$ yields a system of two PDEs for $f\left(R,T\right)$ that can be written in a matrix form as
\begin{equation}\label{eq:dfRT-matrix}
	\begin{pmatrix}
	f_{RR} & f_{RT}\\
	f_{TR} & f_{TT}
	\end{pmatrix}
		\begin{pmatrix}
	R-V_\varphi\\
	T-V_\psi
	\end{pmatrix}
	=0.
\end{equation}
This equation is always satisfied for any arbitrary $f(R,T)$ given that the equations of motion for the scalar fields, viz.  $V_\varphi=R$ and $V_\psi=T$, guarantee that the second factor on the left-hand side of Eq.~\eqref{eq:dfRT-matrix} is always zero. One can thus take the equations $V_\varphi=R$ and $V_\psi=T$, insert the definitions $\varphi=f_R$ and $\psi=f_T$, and integrate these two equations independently to obtain the general form of $f\left(R,T\right)$. Finally, as a verification, the resultant form of $f\left(R,T\right)$ should be inserted into Eq.~\eqref{eq:potentialRT}, which could require some extra constraint to be imposed in the form of the function.\par 
 Equation \eqref{eq:solVpsi-psi} for $V_\psi$ can be written in the form
\begin{equation}\label{eq:TfuncfT}
V_\psi =T= (3w-1)\rho_0\left(\frac{f_T}{\psi_0}\right)^{-\frac{2\left(1+w\right)}{\left(1-w\right)}}.
\end{equation}
This equation is undefined at $w=1$. If $w=\{-1,1/3\}$, this equation states that there is no dependence of $f_T$ in $T$, and thus the dependence of $f\left(R,T\right)$ in $T$ is arbitrary, and, as it stands, any well-behaved function $f\left(R,T\right)$ will be a solution of the partial differential equation. For all other cases, when $w\neq\{-1,1/3,1\}$, Eq.~\eqref{eq:TfuncfT} can be inverted to write $f_T\left(T\right)$ and subsequently integrated to obtain $f\left(R,T\right)$. Since this equation does not depend explicitly on $R$, this implies that the function $f\left(R,T\right)$ is separable in $R$ and $T$, i.e., one can write ${f\left(R,T\right)=f_0+f_1\left(R\right)+f_2\left(T\right)}$, where $f_0$ is a constant, and $f_1(R)$ and $f_2(T)$ are arbitrary functions of their respective argument. Thus, after integrating $f_T\left(T\right)$ one obtains
\begin{equation}\label{eq:f2T}
f_2(T)=\psi_0 \frac{2\left(1+w\right)}{\left(1+3w\right)}  \left[\left(3w-1\right)\rho_0\right]^{\frac{\left(1-w\right)}{2\left(1+w\right)}}
T^{\frac{\left(1+3w\right)}{2\left(1+w\right)}},
\end{equation}
which is valid for $w\neq \left\{-1,1/3,1\right\}$. When $w=0$, the other case studied in this work, Eq.~\eqref{eq:f2T} gives ${f(R,T)= f_1(R)\pm 2\psi_0\sqrt{-\rho_0 T}}$, where the $\pm$ sign was introduced due to taking the square root in the step from Eq.~\eqref{eq:TfuncfT} to Eq.~\eqref{eq:f2T} in the $w=0$ case. The argument inside the square root is positive, since when $w=0$ it is true that $T=-\rho\leq 0$.

Let us now turn to the dependence in $R$. For two of the particular cases studied in this work, $V_\varphi$ was a constant, i.e., with no explicit dependence in $f_R$. In this case, a similar argument to the case where Eq.~\eqref{eq:TfuncfT} does not depend on $f_T$ applies: because there is not an explicit relation between $f_R$ and $R$, there are no constraints on the dependence of $f\left(R,T\right)$ in $R$. In other words, the function $f\left(R,T\right)$ can be left arbitrary or, if Eq.~\eqref{eq:f2T} applies, $f_1\left(R\right)$ can be left arbitrary. The only exception analyzed in this work in which $V_\varphi$ was not a constant was the case when $a\left(t\right)$ is given by Eq.~\eqref{eq:a-m}, viz. ${a\propto t^{2/3}}$. In this case, even with $k=0$, we found that $V_\varphi$ depends on $t$ and therefore on $\varphi$. A potential found analytically with $w=0$ is in Eq.~\eqref{eq:solV-mk0w0-phi1phi20}, from where we see that $V_\varphi$ has an explicit dependence in $\varphi$. Thus $f_R$ has an explicit dependence in $R$. In such cases, one can invert these relations to obtain $f_R\left(R\right)$ which, since it does not depend explicitly on $T$, also requires $f\left(R,T\right)$ to be separable. To obtain the explicit dependence of $f_1\left(R\right)$ in $R$ we integrate $f_R\left(R\right)$ with respect to $R$. With the potential of Eq.~\eqref{eq:solV-mk0w0-phi1phi20}, we find the two following possible solutions (which differ by a sign): 
\begin{equation}\label{eq:f1R}
f_1\left(R\right)=6\pi\rho_0 t_0^2R \pm \frac{2\rho_0\psi_0 t_0}{\sqrt{3}}\sqrt{R}.
\end{equation}
Thus, taking ${f\left(R,T\right)=f_0+f_1\left(R\right)+f_2\left(T\right)}$, with $f_1(R)$ given by Eq.~\eqref{eq:f1R} and $f_2(T)$ given by Eq.~\eqref{eq:f2T}, and using it in Eq.~\eqref{eq:potentialRT} together with the potential in Eq.~\eqref{eq:solV-mk0w0-phi1phi20} one obtains ${f_0=-V_0}$. Therefore,
\begin{equation}
f\left(R,T\right)=-V_0+6\pi\rho_0 t_0^2R \pm \frac{2\rho_0\psi_0 t_0}{\sqrt{3}}\sqrt{R} \pm 2\psi_0\sqrt{-\rho_0 T}.
\end{equation}
is consistent with the particular case of a scale factor evolution that follows a power law of the form given in Eq.~\eqref{eq:a-m}, in a flat geometry (${k=0}$) and with a dust-like fluid (${w=0}$).

\section{Discussion and Conclusions}\label{sec:conclusion}

In this work, we have analyzed the scalar-tensor representation of $f(R,T)$ modified gravity, which includes two scalar fields. This representation provides a simpler framework for the study of cosmology since it results at most in second order differential equations. We preformed our analysis for the case of a perfect fluid in a FLRW universe. We have required the energy-momentum of the fluid to be conserved, which is not a necessary condition in $f\left(R,T\right)$ gravity while it is automatically assured in GR. In order to solve the system of equations, using reconstruction methods,  we have chosen different forms of the scale factor $a(t)$ currently thought to describe well the evolution history of the universe. Even then, the curvature parameter $k$ and equation of state $w$ are still free parameters. This is in contrast to what happens in GR where, for instance, to each equation of state corresponds a unique evolution $a(t)$. 

We found solutions in flat geometry with generic $w$ as well as particular solutions setting the equation of state to the commonly used values.  In the solutions found, the scalar fields $\varphi(t)$ and $\psi(t)$ tend to dominate over time. The field $\psi$ is independent of the curvature parameter $k$ and tends to be associated with the matter content, which is understandable by its definition being related to the trace of the stress-energy tensor $T$. On the other hand, the solutions for the field $\varphi$, whose definition is related to the curvature scalar $R$, depends on the curvature parameter. However it is difficult to interpret these fields, since the two scalar fields $\varphi$ and $\psi$ are introduced as auxiliary fields and as such might not correspond to physical entities. Nevertheless, we understand that these extra gravitational components may act as effective dark energy allowing extra degrees of freedom, such that, for instance, a de Sitter solution can correspond to a range of values of the equation of state, even that of matter or radiation. 
Furthermore, we have attempted to find explicit forms of the function $f(R,T)$ for the cases studied, however this was not always possible, given that the equations found are not always invertible. Thus, in most cases under consideration, any well behaved function $f(R,T)$ would be consistent. In the cases we were able to obtain an explicit form, we have found $f(R,T)$ to be separable.

We have also found the solutions for the scalar field $\varphi$ and the potential $V$ to be discontinuous in terms of the equation of state parameter $w$ for some particular cases, with these discontinuities arising from the use of inverse functions in our methods. We note that these discontinuities are not problematic in the search of a realistic cosmological behavior, since in the context of modified gravity one can always assume the equation of state parameter to be a constant throughout the entire time evolution and the changes in the behavior of the scalar factor to be controlled by the extra scalar degrees of freedom of the gravitational sector, contrary for GR in which different cosmological behaviors are associated with different matter distributions.

In this work, we have chosen different forms of the scale factor from each of the cosmological eras, so it would be an interesting next step to see if it is possible to find solutions that are simultaneously consistent with all epochs of the expansion. Other possibilities for further study of this theory might include exploring the non-conservation of matter, analyzing whether perturbation dependent observables can distinguish $f(R,T)$ from GR, and using a novel alternative to reconstruction methods developed as a generic dynamical system formulation for $f(R)$ gravity \cite{Chakraborty:2021jku}. In particular, the dynamical system approach to analyze the cosmological phase space of the theory in the scalar-tensor representation could provide important insights on the stability of the solutions presented, i.e., if the solutions found correspond to attractors in the phase space of the theory, since the methodology applied in this work does not allow us to trace conclusions in this regard. We are now working on this latter issue, as well as analyzing whether sudden singularities might appear in a finite time in this theory. We hope to report new results on this topics in a near future.

\begin{acknowledgments}

This work was supported by Funda\c{c}\~{a}o para a Ci\^{e}ncia e a Tecnologia  (FCT) through the research grants No. UIDB/04434/2020 and No. UIDP/04434/2020.
T.B.G. acknowledges support from a Ph.D. Research Fellowship in the context of the Funda\c{c}\~{a}o para a Ci\^{e}ncia e a Tecnologia  (FCT) project ``DarkRipple'' with reference PTDC/FIS-OUT/29048/2017.
J.L.R. was supported by the European Regional Development Fund and the programme Mobilitas Pluss (MOBJD647).
F.S.N.L. acknowledges support from the Funda\c{c}\~{a}o para a Ci\^{e}ncia e a Tecnologia (FCT) Scientific Employment Stimulus contract with reference CEECINST/00032/2018, and funding from the research grants No. PTDC/FIS-OUT/29048/2017 and No. CERN/FIS-PAR/0037/2019. 
\end{acknowledgments}




\begin{thebibliography}{99}

\bibitem{Nojiri:2006ri}
S.~Nojiri and S.~D.~Odintsov,
``Introduction to modified gravity and gravitational alternative for dark energy,''
eConf \textbf{C0602061}, 06 (2006)
[arXiv:hep-th/0601213 [hep-th]].

\bibitem{Lobo:2008sg}
F.~S.~N.~Lobo,
``The Dark side of gravity: Modified theories of gravity,''
[arXiv:0807.1640 [gr-qc]].

\bibitem{Nojiri:2010wj}
S.~Nojiri and S.~D.~Odintsov,
``Unified cosmic history in modified gravity: from $F(R)$ theory to Lorentz non-invariant models,''
Phys. Rept. \textbf{505}, 59-144 (2011)
[arXiv:1011.0544 [gr-qc]].

\bibitem{Clifton:2011jh}
T.~Clifton, P.~G.~Ferreira, A.~Padilla and C.~Skordis,
``Modified Gravity and Cosmology,''
Phys. Rept. \textbf{513}, 1-189 (2012)
[arXiv:1106.2476 [astro-ph.CO]].

\bibitem{Capozziello:2011et}
S.~Capozziello and M.~De Laurentis,
``Extended Theories of Gravity,''
Phys. Rept. \textbf{509}, 167-321 (2011)
[arXiv:1108.6266 [gr-qc]].

\bibitem{CANTATA:2021ktz}
E.~N.~Saridakis \textit{et al.} [CANTATA],
``Modified Gravity and Cosmology: An Update by the CANTATA Network,''
[arXiv:2105.12582 [gr-qc]].

\bibitem{Avelino:2016lpj}
P.~Avelino, T.~Barreiro, C.~S.~Carvalho, A.~da Silva, F.~S.~N.~Lobo, P.~Martin-Moruno, J.~P.~Mimoso, N.~J.~Nunes, D.~Rubiera-Garcia and D.~Saez-Gomez, \textit{et al.}
``Unveiling the Dynamics of the Universe,''
Symmetry \textbf{8}, no.8, 70 (2016)
[arXiv:1607.02979 [astro-ph.CO]].

\bibitem{SupernovaCosmologyProject:1998vns}
S.~Perlmutter \textit{et al.} [Supernova Cosmology Project],
``Measurements of $\Omega$ and $\Lambda$ from 42 high redshift supernovae,''
Astrophys. J. \textbf{517}, 565-586 (1999)
[arXiv:astro-ph/9812133 [astro-ph]].

\bibitem{SupernovaSearchTeam:1998fmf}
A.~G.~Riess \textit{et al.} [Supernova Search Team],
``Observational evidence from supernovae for an accelerating universe and a cosmological constant,''
Astron. J. \textbf{116}, 1009-1038 (1998)
[arXiv:astro-ph/9805201 [astro-ph]].

\bibitem{Sotiriou:2008rp}
T.~P.~Sotiriou and V.~Faraoni,
``$f(R)$ Theories Of Gravity,''
Rev. Mod. Phys. \textbf{82}, 451-497 (2010)
[arXiv:0805.1726 [gr-qc]].

\bibitem{Capozziello:2002rd}
S.~Capozziello,
``Curvature quintessence,''
Int. J. Mod. Phys. D \textbf{11}, 483-492 (2002)
[arXiv:gr-qc/0201033 [gr-qc]].

\bibitem{Olmo:2011uz}
G.~J.~Olmo,
``Palatini Approach to Modified Gravity: $f(R)$ Theories and Beyond,''
Int. J. Mod. Phys. D \textbf{20}, 413-462 (2011)
[arXiv:1101.3864 [gr-qc]].

\bibitem{Harko:2011nh}
T.~Harko, T.~S.~Koivisto, F.~S.~N.~Lobo and G.~J.~Olmo,
``Metric-Palatini gravity unifying local constraints and late-time cosmic acceleration,''
Phys. Rev. D \textbf{85}, 084016 (2012)
[arXiv:1110.1049 [gr-qc]].

\bibitem{Harko:2020ibn}
T.~Harko and F.~S.~N.~Lobo,
``Beyond Einstein's General Relativity: Hybrid metric-Palatini gravity and curvature-matter couplings,''
Int. J. Mod. Phys. D \textbf{29}, no.13, 2030008 (2020)
[arXiv:2007.15345 [gr-qc]].


\bibitem{Harko:2018ayt}
T.~Harko and F.~S.~N.~Lobo,
{\it{Extensions of f(R) Gravity: Curvature-Matter Couplings and Hybrid Metric Palatini Theory}}, Cambridge University Press, Cambridge, (2018).

\bibitem{Capozziello:2012ny}
S.~Capozziello, T.~Harko, T.~S.~Koivisto, F.~S.~N.~Lobo and G.~J.~Olmo,
``Cosmology of hybrid metric-Palatini $f(X)$-gravity,''
JCAP \textbf{04}, 011 (2013)
[arXiv:1209.2895 [gr-qc]].

\bibitem{Capozziello:2013uya}
S.~Capozziello, T.~Harko, F.~S.~N.~Lobo and G.~J.~Olmo,
``Hybrid modified gravity unifying local tests, galactic dynamics and late-time cosmic acceleration,''
Int. J. Mod. Phys. D \textbf{22}, 1342006 (2013)
[arXiv:1305.3756 [gr-qc]].

\bibitem{Capozziello:2015lza}
S.~Capozziello, T.~Harko, T.~S.~Koivisto, F.~S.~N.~Lobo and G.~J.~Olmo,
``Hybrid metric-Palatini gravity,''
Universe \textbf{1}, no.2, 199-238 (2015)
[arXiv:1508.04641 [gr-qc]].

\bibitem{Rosa:2017jld}
J.~L.~Rosa, S.~Carloni, J.~P.~d.~Lemos and F.~S.~N.~Lobo,
``Cosmological solutions in generalized hybrid metric-Palatini gravity,''
Phys. Rev. D \textbf{95} (2017) no.12, 124035
doi:10.1103/PhysRevD.95.124035
[arXiv:1703.03335 [gr-qc]].

\bibitem{Rosa:2019ejh}
J.~L.~Rosa, S.~Carloni and J.~P.~S.~Lemos,
``Cosmological phase space of generalized hybrid metric-Palatini theories of gravity,''
Phys. Rev. D \textbf{101} (2020) no.10, 104056
doi:10.1103/PhysRevD.101.104056
[arXiv:1908.07778 [gr-qc]].

\bibitem{Rosa:2021ish}
J.~L.~Rosa, F.~S.~N.~Lobo and D.~Rubiera-Garcia,
``Sudden singularities in generalized hybrid metric-Palatini cosmologies,''
doi:10.1088/1475-7516/2021/07/009
[arXiv:2103.02580 [gr-qc]].

\bibitem{Harko:2011kv}
T.~Harko, F.~S.~N.~Lobo, S.~Nojiri and S.~D.~Odintsov,
``$f(R,T)$ gravity,''
Phys. Rev. D \textbf{84}, 024020 (2011)
[arXiv:1104.2669 [gr-qc]].

\bibitem{Poplawski:2006ey}
N.~J.~Poplawski,
``A Lagrangian description of interacting dark energy,''
[arXiv:gr-qc/0608031 [gr-qc]].

\bibitem{Jamil:2011ptc}
M.~Jamil, D.~Momeni, M.~Raza and R.~Myrzakulov,
``Reconstruction of some cosmological models in $f(R,T)$ gravity,''
Eur. Phys. J. C \textbf{72}, 1999 (2012)
[arXiv:1107.5807 [physics.gen-ph]].

\bibitem{Houndjo:2011fb}
M.~J.~S.~Houndjo and O.~F.~Piattella,
``Reconstructing $f(R,T)$ gravity from holographic dark energy,''
Int. J. Mod. Phys. D \textbf{21}, 1250024 (2012)
[arXiv:1111.4275 [gr-qc]].

\bibitem{Houndjo:2011tu}
M.~J.~S.~Houndjo,
``Reconstruction of $f(R,T)$ gravity describing matter dominated and accelerated phases,''
Int. J. Mod. Phys. D \textbf{21}, 1250003 (2012)
[arXiv:1107.3887 [astro-ph.CO]].

\bibitem{Alvarenga:2013syu}
F.~G.~Alvarenga, A.~de la Cruz-Dombriz, M.~J.~S.~Houndjo, M.~E.~Rodrigues and D.~S\'aez-G\'omez,
``Dynamics of scalar perturbations in $f(R,T)$ gravity,''
Phys. Rev. D \textbf{87}, no.10, 103526 (2013)
[erratum: Phys. Rev. D \textbf{87}, no.12, 129905 (2013)]
[arXiv:1302.1866 [gr-qc]].

\bibitem{Shabani:2013djy}
H.~Shabani and M.~Farhoudi,
``$f(R,T)$ Cosmological Models in Phase Space,''
Phys. Rev. D \textbf{88}, 044048 (2013)
[arXiv:1306.3164 [gr-qc]].

\bibitem{Shabani:2014xvi}
H.~Shabani and M.~Farhoudi,
``Cosmological and Solar System Consequences of $f(R,T)$ Gravity Models,''
Phys. Rev. D \textbf{90}, no.4, 044031 (2014)
[arXiv:1407.6187 [gr-qc]].

\bibitem{Moraes:2015kka}
P.~H.~R.~S.~Moraes,
``Cosmological solutions from Induced Matter Model applied to 5D $f(R,T)$ gravity and the shrinking of the extra coordinate,''
Eur. Phys. J. C \textbf{75}, no.4, 168 (2015)
[arXiv:1502.02593 [gr-qc]].

\bibitem{Rosa:2021tei}
J.~L.~Rosa, M.~A.~Marques, D.~Bazeia and F.~S.~N.~Lobo,
``Thick branes in the scalar-tensor representation of $f(R,T)$ gravity,''
[arXiv:2105.06101 [gr-qc]].

\bibitem{Rosa:2021myu}
J.~L.~Rosa, D.~Bazeia and A.~S.~Lob\~ao,
``Effects of Cuscuton dynamics on braneworld configurations in the scalar-tensor representation of $f\left(R,T\right)$ gravity,''
[arXiv:2111.08089 [gr-qc]].

\bibitem{Haghani:2013oma}
Z.~Haghani, T.~Harko, F.~S.~N.~Lobo, H.~R.~Sepangi and S.~Shahidi,
``Further matters in space-time geometry: $f(R, T, R_{\mu\nu} T^{\mu\nu})$ gravity,''
Phys. Rev. D \textbf{88}, no.4, 044023 (2013)
[arXiv:1304.5957 [gr-qc]].

\bibitem{Odintsov:2013iba}
S.~D.~Odintsov and D.~S\'aez-G\'omez,
``$f(R, T, R_{\mu\nu} T^{\mu\nu})$ gravity phenomenology and $\Lambda$CDM universe,''
Phys. Lett. B \textbf{725}, 437-444 (2013)
[arXiv:1304.5411 [gr-qc]].


\bibitem{Zaregonbadi:2016xna}
R.~Zaregonbadi, M.~Farhoudi and N.~Riazi,
``Dark Matter From $f(R,T)$ Gravity,''
Phys. Rev. D \textbf{94}, 084052 (2016)
[arXiv:1608.00469 [gr-qc]].

\bibitem{Moraes:2017mir}
P.~H.~R.~S.~Moraes and P.~K.~Sahoo,
``Modelling wormholes in $f(R,T)$ gravity,''
Phys. Rev. D \textbf{96}, no.4, 044038 (2017)
[arXiv:1707.06968 [gr-qc]].

\bibitem{Zubair:2016cde}
M.~Zubair, S.~Waheed and Y.~Ahmad,
``Static spherically symmetric wormholes in $f(R,T)$ gravity,''
Eur. Phys. J. C \textbf{76}, no.8, 444 (2016)
[arXiv:1607.05998 [gr-qc]].

\bibitem{Moraes:2016akv}
P.~H.~R.~S.~Moraes, R.~A.~C.~Correa and R.~V.~Lobato,
``Analytical general solutions for static wormholes in $f(R,T)$ gravity,''
JCAP \textbf{07}, 029 (2017)
[arXiv:1701.01028 [gr-qc]].

\bibitem{Das:2016mxq}
A.~Das, F.~Rahaman, B.~K.~Guha and S.~Ray,
``Compact stars in $f(R,T)$ gravity,''
Eur. Phys. J. C \textbf{76}, no.12, 654 (2016)
[arXiv:1608.00566 [gr-qc]].

\bibitem{Horndeski:1974wa}
G.~W.~Horndeski,
``Second-order scalar-tensor field equations in a four-dimensional space,''
Int. J. Theor. Phys. \textbf{10}, 363-384 (1974)

\bibitem{Deffayet:2011gz}
C.~Deffayet, X.~Gao, D.~A.~Steer and G.~Zahariade,
``From k-essence to generalised Galileons,''
Phys. Rev. D \textbf{84}, 064039 (2011)
[arXiv:1103.3260 [hep-th]].

\bibitem{Rosa:2021teg}
J.~L.~Rosa,
``Junction conditions and thin shells in perfect-fluid $f(R,T)$ gravity,''
Phys. Rev. D \textbf{103}, no.10, 104069 (2021)
[arXiv:2103.11698 [gr-qc]].

\bibitem{Capozziello:2005ku}
S.~Capozziello, V.~F.~Cardone and A.~Troisi,
``Reconciling dark energy models with $f(R)$ theories,''
Phys. Rev. D \textbf{71}, 043503 (2005)
[arXiv:astro-ph/0501426 [astro-ph]].

\bibitem{Multamaki:2005zs}
T.~Multamaki and I.~Vilja,
``Cosmological expansion and the uniqueness of gravitational action,''
Phys. Rev. D \textbf{73}, 024018 (2006)
[arXiv:astro-ph/0506692 [astro-ph]].

\bibitem{Bertolami:2008ab}
O.~Bertolami, F.~S.~N.~Lobo and J.~Paramos,
``Non-minimum coupling of perfect fluids to curvature,''
Phys. Rev. D \textbf{78}, 064036 (2008)
[arXiv:0806.4434 [gr-qc]].

\bibitem{Koivisto:2005yk}
T.~Koivisto,
``Covariant conservation of energy momentum in modified gravities,''
Class. Quant. Grav. \textbf{23}, 4289-4296 (2006)
[arXiv:gr-qc/0505128 [gr-qc]].

\bibitem{Bertolami:2007gv}
O.~Bertolami, C.~G.~Boehmer, T.~Harko and F.~S.~N.~Lobo,
``Extra force in f(R) modified theories of gravity,''
Phys. Rev. D \textbf{75}, 104016 (2007)
[arXiv:0704.1733 [gr-qc]].

\bibitem{Harko:2010mv}
T.~Harko and F.~S.~N.~Lobo,
``$f(R,L_{m})$ gravity,''
Eur. Phys. J. C \textbf{70}, 373-379 (2010)
[arXiv:1008.4193 [gr-qc]].

\bibitem{Harko:2012hm}
T.~Harko, F.~S.~N.~Lobo and O.~Minazzoli,
``Extended $f(R,L_m)$ gravity with generalized scalar field and kinetic term dependences,''
Phys. Rev. D \textbf{87}, no.4, 047501 (2013)
[arXiv:1210.4218 [gr-qc]].

\bibitem{Harko:2014gwa}
T.~Harko and F.~S.~N.~Lobo,
``Generalized curvature-matter couplings in modified gravity,''
Galaxies \textbf{2}, no.3, 410-465 (2014)
[arXiv:1407.2013 [gr-qc]].

\bibitem{Harko:2014sja}
T.~Harko, F.~S.~N.~Lobo, G.~Otalora and E.~N.~Saridakis,
``Nonminimal torsion-matter coupling extension of $f(T)$ gravity,''
Phys. Rev. D \textbf{89}, 124036 (2014)
[arXiv:1404.6212 [gr-qc]].

\bibitem{Harko:2014aja}
T.~Harko, F.~S.~N.~Lobo, G.~Otalora and E.~N.~Saridakis,
``$f(T,\mathcal{T})$ gravity and cosmology,''
JCAP \textbf{12}, 021 (2014)
[arXiv:1405.0519 [gr-qc]].

\bibitem{Harko:2018gxr}
T.~Harko, T.~S.~Koivisto, F.~S.~N.~Lobo, G.~J.~Olmo and D.~Rubiera-Garcia,
``Coupling matter in modified $Q$ gravity,''
Phys. Rev. D \textbf{98}, no.8, 084043 (2018)
[arXiv:1806.10437 [gr-qc]].

\bibitem{Harko:2014pqa}
T.~Harko,
``Thermodynamic interpretation of the generalized gravity models with geometry-matter coupling,''
Phys. Rev. D \textbf{90}, no.4, 044067 (2014)
[arXiv:1408.3465 [gr-qc]].

\bibitem{Harko:2015pma}
T.~Harko, F.~S.~N.~Lobo, J.~P.~Mimoso and D.~Pav\'on,
``Gravitational induced particle production through a nonminimal curvature-matter coupling,''
Eur. Phys. J. C \textbf{75}, 386 (2015)
[arXiv:1508.02511 [gr-qc]].

\bibitem{Harko:2021bdi}
T.~Harko, F.~S.~N.~Lobo and E.~N.~Saridakis,
``Gravitationally Induced Particle Production through a Nonminimal Torsion-Matter Coupling,''
Universe \textbf{7}, no.7, 227 (2021)
[arXiv:2107.01937 [gr-qc]].

\bibitem{Chakraborty:2021jku}
S.~Chakraborty, K.~MacDevette and P.~Dunsby,
``A model independent approach to the study of $f(R)$ cosmologies with expansion histories close to $\Lambda$CDM,''
Phys. Rev. D \textbf{103}, no.12, 124040 (2021)
[arXiv:2103.02274 [gr-qc]].

\end{thebibliography}
\end{document}